%% file: thesis.tex
\documentclass[12pt,twoside]{report} 
\usepackage{graphicx} 
\usepackage{makeidx} 
\usepackage{amsmath} 
\usepackage{amsthm}
\usepackage[bookmarks=true]{hyperref} 
\usepackage{color}
\usepackage{times}
\usepackage{amssymb}
\usepackage{euler}
\usepackage{subfigure}
\makeindex 

\newcommand{\er}[1]{\mathbb{#1}}
\newcommand{\logic}[1]{\mathfrak{#1}}
\newcommand{\eLOQC}{efficient linear optics quantum computation}
\newcommand{\ket}[1]{\left|{#1}\right\rangle}
\newcommand{\bra}[1]{\left\langle{#1}\right|}
\newcommand{\braket}[2]{\left\langle{#1}|{#2}\right\rangle}
\newcommand{\enc}[1]{\overline{#1}}
\newcommand{\CNOT}{{\tt CNOT}}
\newcommand{\CSIGN}{{\tt CSIGN}}
\newcommand{\NS}{{\tt NS}}
\newcommand{\SWAP}{{\tt SWAP}}
\newcommand{\ox}{\otimes}
\newcommand{\X}{{\bf X}}
\newcommand{\Y}{{\bf Y}}
\newcommand{\Z}{{\bf Z}}
\newcommand{\I}{{\bf I}}
\newcommand{\Had}{{\bf H}}
\newcommand{\Pha}{{\bf P}}

\newcommand{\op}[1]{{\bf #1}}
\newcommand{\Pauli}{{\mathcal P}}
\newcommand{\R}{\equiv}
\newcommand{\C}{{\mathscr C}}
\newcommand{\etal}{{\em et al.}}
\newcommand{\e}{\epsilon}
\newcommand{\dt}{\delta}
\newcommand{\degg}{{\circ}}

\newtheorem{definition}{Definition}

\newtheorem{proposition}{Proposition}
\newtheorem{theorem}{Theorem}

 {\begin{list}{}%
  {%
   \settowidth{\labelwidth}{\textnormal{#1 =}}%
   \setlength{\leftmargin}{\labelwidth}%
   \addtolength{\leftmargin}{\labelsep}%
   \setlength{\itemsep}{-\parsep}}}%
 {\end{list}}

\begin{document}
\include{thesis_frontpages} 

\pagestyle{myheadings}
\markboth{Erasure Thresholds for Linear Optics}
\normalsize

\include{Intro}
\include{Qecc-ft}

\include{eLOQC}

\include{Candidate-Codes}
\include{Threshold}

\include{Simulation}

\include{Conclusion}

\appendix
\include{calc-z}

\include{calc-full}


\bibliographystyle{hplain} 
\bibliography{thesis-bibliography} 

\printindex 
\end{document}

%% file: thesis_frontpages.tex

\pagestyle{empty} 

\begin{center}

\vspace*{1.0cm}
\Huge
{\bf Erasure Thresholds for \\
Efficient Linear Optics Quantum Computation }

\vspace*{1.0cm}
\normalsize
by \\
\vspace*{1.0cm}
\Large
Marcus Palmer da Silva\\
\vspace*{2.0cm}
\normalsize
A thesis \\
presented to the University of Waterloo \\ 
in fulfillment of the \\
thesis requirement for the Master's degree \\
in \\
Physics\\

\vspace*{2.0cm}
Waterloo, Ontario, December 2003\\
(Revised May 2004)

\vspace*{1.0cm}
\copyright Marcus Palmer da Silva, 2003, 2004\\

\end{center}

\newpage
 

\pagestyle{plain} 
\pagenumbering{roman} 
\setcounter{page}{2}

\noindent
I hereby declare that I am the sole author of this thesis.

\noindent
I authorize the University of Waterloo to lend this thesis to other
institutions or individuals for the purpose of scholarly research.
\vspace{4cm}

\noindent
Marcus Silva

\vspace{4cm}

\noindent
I authorize the University of Waterloo to reproduce this thesis by
photocopying or other means, in total or in part, at the request of other
institutions or individuals for the purpose of scholarly research.
\vspace{4cm}

\noindent
Marcus Silva
\newpage

\noindent
The University of Waterloo requires the signatures of all persons using
or photocopying this thesis. Please sign below, and give address and date.
\newpage

\begin{center}\textbf{Acknowledgments}\end{center}

First and foremost, I thank my parents, Luis and  Carol, for everything.

Working at the IQC in Waterloo has been a great opportunity, and I have
made many friends from whom I have learned a lot. I must thank
Michele Mosca, Christof Zalka, and Martin R\"otteler for their support,
patience, time, encouragement, and insight; Raymond Laflamme for 
the encouragement and support; and Wendy Reibel for all the help.
I also thank Daniel Gottesman and Achim Kempf, for their interest in
my research and their kind agreement to serve on my committee, along with
Michele Mosca, Christof Zalka, Raymond Laflamme and Martin R\"otteler. 

My close friends Laura, Robbi, Ben, my distant friends Ying and Lu, and
all other friends in between who helped me unwind when I needed to, should
be thanking me for being their friend. 

Most importantly, I thank Wenting, for her patience.

\newpage

\begin{center}
\Large
\textbf{Abstract}
\end{center}
Using an error models motivated by the Knill, Laflamme,
Milburn proposal for efficient linear optics quantum computing [Nature {\bf 409},46--52, 2001],
error rate thresholds for erasure errors caused by imperfect photon detectors
using a 7 qubit code are derived and
verified through simulation. A novel method -- based on a Markov chain
description of the erasure correction procedure -- is developed and used
to calculate the recursion relation describing the error rate at
different encoding levels from which the threshold is derived, matching
threshold predictions by Knill, Laflamme and Milburn [quant-ph/0006120, 2000].
In particular, the erasure threshold for gate failure rate in the same
order as the measurement failure rate is found to be above $1.78\%$.
\newpage

\setcounter{page}{6} 
\tableofcontents
\listoffigures
\newpage

\pagenumbering{arabic}

%% file: Intro.tex
\chapter{Introduction}
\markright{Introduction}
Quantum computation was born when Benioff \cite{benioff} proposed
using the laws of quantum mechanics instead of classical mechanics 
to perform computation.
More importantly for physicists, Feynman \cite{feynman}
pointed out, was the possibility of simulating quantum system by using
such quantum computers -- because the state space of quantum systems
has dimensions that grow exponentially with the number of subsystems
involved, they are notoriously inefficient to simulate with classical
computers.

Much progress has been done on how to perform certain algorithms
much more efficiently in quantum computers, as well as how to simulate
quantum systems with quantum computers. The main challenge is to
construct physical systems on which such computers may be built. Many 
proposals have been put forth, each with its strengths and weaknesses,
but the recent proposal by Knill, Laflamme and Milburn \cite{klm:2001}, which uses
only linear optics elements and single photon sources and detectors,
is of particular interest for various reasons. First of all, it was
believed that it was impossible to build a universal quantum computer 
from linear optics elements because photons do not interact with
each other. This proposal relies heavily on state preparation in order
to avoid such a hurdle. Moreover, one of the most successful applications
of quantum computing, quantum key distribution, would benefit directly
from the construction of an optical quantum computer because most
protocols are implemented with optical communication. Quantum computation
with non-linear optics, on the other hand, is notorious for the
photon loss, making it extremely inefficient.

The main problem in a physical implementation of a quantum computer
is its sensitivity to error and to the environment. In a classical
computer, information is represented by two different states, $\logic{0}$
and $\logic{1}$, and such computers can easily be built so that error in
distinguishing between these two states is insignificant. 
In a quantum computer, the state may be in a
superposition of these two basis states, say $\alpha\ket{\logic{0}}+\beta\ket{\logic{1}}$,
and when the state is measured, we obtain $\ket{\logic{0}}$ with probability $|\alpha|^2$,
and $\ket{\logic{1}}$ with probability $|\beta|^2$. The information
in a quantum computer ends up being more akin to an analog data than
to digital data, even though there is only
a discrete number of basis states. Imprecision in devices used for
computation, the logic gates, has a much greater impact in quantum
information than in classical information. Moreover, it is much
harder to isolate a quantum system from the environment, so that
interaction between the two also end up introducing error into
the computation, in an effect known as decoherence, which is
only observed in quantum systems.

The linear optics quantum computing proposal also suffers from these problems,
even though photons are much less sensitive to decoherence than other
physical systems proposed for quantum computation. The reason is that
the gates proposed for efficient linear optics quantum computation 
are probabilistic, so that even with infinite precision in
the linear optics elements, they may fail. The 
authors of the proposal have demonstrated that by encoding the
data carefully, one can easily overcome the probabilistic nature of
the gates \cite{klm-thr:2000}. However, these gates depend on 
single photon measurements, which are known to have low efficiency, although the gates
described in the proposal can flag when detectors have failed and replace
the lost qubit automatically. The natural question then is what is the maximum rate
at which photons may be lost in these gates while still allowing for
scalable quantum computation? The main objective of this thesis is
to answer that question.

This thesis is organized as follows. Chapter 2 gives a basic overview
of quantum erasure correction codes and fault-tolerant computation, focusing
on CSS codes and the stabilizer formalism. Chapter 3 describes the basics
of the efficient linear optics quantum computation proposal by Knill, Laflamme and Milburn,
focusing on the description of the error model for teleportation failures due
to the inherent probabilistic nature of the linear optics
implementation, as well as photon loss due to detector inefficiencies.
Chapter 4 gives a detailed description of the Steane code, and contrasts it
to the Grassl code in the context of the error model for linear optics
quantum computing.
Given these background chapters, it becomes clear how to 
perform the erasure correction procedure, and in Chapter 5 a new and compact
description of the procedure is given in terms of Markov chains.
With this description the threshold is obtained
for both error models, with the detailed calculations being presented
in the appendices. In Chapter 6 a Monte Carlo simulation of the erasure
correction procedure is described, and the simulation results are presented and
contrasted with the theoretical prediction of Chapter 5. The thesis concludes
with a discussion of the results, and suggestions for future work.
\section{Notation}
Some basic notation is assumed in this thesis, and it is briefly
reviewed here. Other notation is introduced in the body of the thesis
as needed, along with explanations.

In order to emphasize the difference between qubits and photon number states,
qubits in the computational basis will be represented with gothic font,
e.g. $\ket{\logic{0}}$ and $\ket{\logic{1}}$, while photon numbers will be
represented in the usual fonts, e.g. $\ket{1}$ and $\ket{2}$.
Operators acting on the physical qubits (qubits not encoded by
error correction code) are denoted by bold 
capital letters, e.g. $\X$, and multi-letter operators use 
typewriter font, e.g. $\CNOT$. Operators acting on the {\em encoded} qubits
are denoted in the same way, with an added overbar, e.g. $\enc{\X}$, and encoded
qubits also have an overbar, e.g. $\ket{\enc{\Psi}}$ or $\ket{\enc{\logic{01}}}$.
If it is clear from the context that the operator is encoded, the overbar is
omitted.

The single qubit Pauli group $\Pauli_1\equiv\Pauli$ consists of the operators
\begin{eqnarray}
\I &=&\begin{pmatrix}1&0\\0&1\end{pmatrix}\\
\X &=&\begin{pmatrix}0&1\\1&0\end{pmatrix}\\ 
\Y &=&\begin{pmatrix}0&-1\\1&0\end{pmatrix}\\ 
\Z &=&\begin{pmatrix}1&0\\0&-1\end{pmatrix},
\end{eqnarray}
along with multiplication by $i=\sqrt{-1}$,
 following the convention in \cite{gottesman:1998}. Under this definition,
products of Pauli operators are given by
\begin{eqnarray}
\X^2&=&\I\\
\Z^2&=&\I\\
\Y^2&=&-\I\\
\X\Z&=&\Y\label.
\end{eqnarray}
The Pauli group $\Pauli_n$ over $n$ qubits is given by the $n$ fold tensor
product of single qubit Pauli operators along with multiplication by $\{\pm i\}$.

Other common unitary operations are
\begin{eqnarray}
\Had &=& \frac{1}{\sqrt{2}} \begin{pmatrix}1&1\\1&-1\end{pmatrix}\\
\Pha &=& \begin{pmatrix}1&0\\0&i\end{pmatrix}\\
\CNOT_{1,2} &=& \begin{pmatrix}1&0&0&0\\0&1&0&0\\0&0&0&1\\0&0&1&0\end{pmatrix}\\
\CSIGN_{1,2} &=& \begin{pmatrix}1&0&0&0\\0&1&0&0\\0&0&1&0\\0&0&0&-1\end{pmatrix},
\end{eqnarray}
where $\Had$ is the Hadamard transform, $\Pha$ is the phase gate, $\CNOT_{1,2}$ is the
controlled-$\X$ where qubit $1$, the control, determines the application of an $\X$ on qubit $2$,
the target, and $\CSIGN_{1,2}$ is the controlled-$\Z$ with control and target similarly defined. Throughout the thesis, $\Had$ and $\Pha$ are taken to be a single qubit operations.

These unitary operations can also be represented by rotations about
Pauli operators. In general, given some operator $\op{U}\in\Pauli_n$,
we define
\begin{equation}
(\op{U})_{\theta}=e^{-i\pi\frac{\theta}{360^{\degg}}\op{U}},
\end{equation}
where $\theta$ is given in degrees. In particular, if $\theta=\pm 180^{\degg}$,
then the rotations are equivalent to operators in $\Pauli_n$, and
if $\theta=\pm 90^\degg$ the rotation is equivalent to a product of
$\Had$, $\Pha$ and $\CNOT$ applied to various different qubits.

%% file: Qecc-ft.tex
\chapter{Erasure Correction and Fault-Tolerance\label{ch:qecc-ft}}
\markright{Erasure Correction and Fault-Tolerance}

A brief overview of erasure errors is given, along with a brief
discussion on the basics of quantum error correction codes and
fault-tolerant computation.
In this discussion, quantum error correction codes are referred to as
{\em error correction codes}. Error correction codes over classical data
will be referred to as {\em classical error correction codes}.

\section{The Erasure Channel\label{sec:erasure-channel}}
A common error model for quantum data is given by a channel for
which there is a finite probability $\Pr(\er{O})$ 
of an error superoperator $\er{O}$
being applied to the qubit transmitted. In that case, we
can define the channel by the superoperator 
\begin{equation}\label{eqn:error-supop}
{\mathcal E}_{\er{O}}(\rho)=(1-\Pr(\er{O}))\rho + \Pr(\er{O})\er{O}(\rho),
\end{equation}
where $\rho$ is the density matrix of the qubit input into the channel,
and we take this channel to be {\em memoryless}, that is, different uses of
the same channel are statistically independent.

In general the corruption of data is not {\em a priori} obvious to the
observer, and as was described in the introduction, 
one must encode the data in
special ways in order to detect such corruption. Under
some physical models, however, it is immediately known when
some error superoperators have been applied.
The canonical example of this
is spontaneous emission in qubits represented by atoms \cite{grass-etal:1997},
where one may detect the resulting photons and determine that the state
of the atom has been corrupted. In general, it is possible to consider the qubit encoding to be a 
Hilbert space strictly smaller than the Hilbert space describing
the entire physical system. Say for example, the computational Hilbert
space is given by 
\begin{equation}
{\mathcal H}_{q}=\text{span}\{\ket{\logic{0}},\ket{\logic{1}}\},
\end{equation}
but the state of the entire physical system is in the Hilbert space 
\begin{equation}
{\mathcal H}=\text{span}\{\ket{\logic{0}},\ket{\logic{1}},\ket{\logic{2}},\cdots,\ket{\logic{n}}\}.
\end{equation}
Errors that map qubits into the space orthogonal to ${\mathcal H}_q$,
in a process known as {\em leakage},
can always be detected without disturbing the computational subspace, and
therefore can be considered erasures, as long as the qubits
determined to be in ${\mathcal H}_q^{\perp}$ are replaced by fresh computational qubits.
This is, in essence, the case for the
linear optics proposal described in Chapter \ref{ch:eloqc},
and it is the main motivation for this thesis.

Abstracting from the implementation details, we can think of 
an erasure channel as an error channel with {\em side information}
through which the application of some error superoperator is flagged, and 
therefore it is known which qubits have been corrupted.
In that case,
because there is certainty that the error superoperator has been
applied, we call it an {\em erasure superoperator}, and we
represent it by a hollowed letter corresponding to the
error superoperator applied, or $\er{O}$.
The important distinction being that it is known with certainty that
the operator has been applied, so the probability of the error $\er{O}$
occurring is absent from this description. 

The superoperator equivalent to a full qubit erasure, or a complete randomization
of the state, is
\begin{equation}\label{full-er-eqn}
\er{E}(\rho)=\I=\frac{1}{4}\left(\rho+\X\rho\X+\Y\rho\Y+\Z\rho\Z\right).
\end{equation}
The partial erasure superoperators, which correspond to partial
randomizations of the qubit (a unique quantum feature) are
given by 
\begin{subequations}\label{partial-er-eqns}
\begin{align}
\er{Z}(\rho)=&\frac{1}{2}\left(\rho+\Z\rho\Z\right)\\
\er{X}(\rho)=&\frac{1}{2}\left(\rho+\X\rho\X\right),
\end{align}
\end{subequations}
corresponding to a possible $\X$ error or a possible $\Z$ error respectively,
and called $\X$ erasure and $\Z$, or {\em phase}, erasure. As will
be demonstrated in Chapter \ref{ch:eloqc}, photon loss due to detector
inefficiencies within a $\CSIGN$s in
linear optics quantum computing can be modeled as a gate application
followed by transmission over an erasure channel.

For the rest of this thesis, we will refer to erasure superoperators
that act independently on multiple qubits as {\em erasure patterns}.
For example, a block of seven qubits where the first 
is affected by a $\Z$ erasure, while the last is affected by a
full erasure, is said to have been affected by the erasure pattern
\begin{equation}\label{eqn:erasure-pat-eg}
\er{Z}\ox\I\ox\I\ox\I\ox\I\ox\I\ox\er{E}.
\end{equation}
This, in effect, describes a convex sum\footnote{In this context,
we take a convex sum to mean the application of
different Pauli operators $\op{O}_i$ with probability
$\Pr(\op{O}_i)\ge 0$ to a state described by some density matrix 
$\rho$ so that $\sum_i\Pr(\op{O}_i)=1$, yielding some
superoperator $$\mathcal{O}(\rho) = \sum_i \Pr(\op{O}_i)\op{O}_i\rho\op{O}_i^\dagger.$$ It is clear that \eqref{full-er-eqn}
and \eqref{partial-er-eqns} fit into this category.} 
of Pauli operators acting on
seven qubits. These Pauli operators are known as {\em error operators},
and they represent the actual errors that are detected by the error
correcting procedure. In our discussion, we will call them
{\em erasure operators} when it is known which qubits {\em could}
have been corrupted.
In particular, given \eqref{eqn:erasure-pat-eg},
measurements may indicate that any erasure operator among
\begin{equation}
\begin{array}{ccccccccccccc}
\I&\ox&\I&\ox&\I&\ox&\I&\ox&\I&\ox&\I&\ox&\I\\
\I&\ox&\I&\ox&\I&\ox&\I&\ox&\I&\ox&\I&\ox&\X\\
\I&\ox&\I&\ox&\I&\ox&\I&\ox&\I&\ox&\I&\ox&\Y\\
\I&\ox&\I&\ox&\I&\ox&\I&\ox&\I&\ox&\I&\ox&\Z\\
\Z&\ox&\I&\ox&\I&\ox&\I&\ox&\I&\ox&\I&\ox&\I\\
\Z&\ox&\I&\ox&\I&\ox&\I&\ox&\I&\ox&\I&\ox&\X\\
\Z&\ox&\I&\ox&\I&\ox&\I&\ox&\I&\ox&\I&\ox&\Y\\
\Z&\ox&\I&\ox&\I&\ox&\I&\ox&\I&\ox&\I&\ox&\Z,
\end{array}
\end{equation}
was applied, and given the types of erasures considered here,
the probability distribution for any of these erasure operators being measured
is always uniform. By definition, the weight of an $n$ qubit Pauli operator
is the number of qubits over which it acts non-trivially.
Similarly, the number of qubits over which some erasure pattern $\er{W}$
{\em may} act non-trivially is called the {\em weight} of the pattern, denoted $wt(\er{W})$.
\section{Conditions for Quantum Erasure Codes}
There are special conditions for an encoding of quantum data to be
considered a quantum error correcting code, and they hinge on
what error operators a code claims to be able to correct. The
conditions developed by Knill and Laflamme \cite{kl-cond:1997} are,
given a quantum code consisting of 
encoded states $\left\{\ket{c_i}\right\}_{i=1}^k$ and correcting
a set of error operators $\{\op{E}_j\}$ 
\begin{subequations}
\begin{eqnarray}
\bra{c_l}\op{E}_i^\dagger \op{E}_j\ket{c_l}&=&\bra{c_m}\op{E}_i^\dagger \op{E}_j\ket{c_m}\label{eqn:kl-c1}\\
\bra{c_l}\op{E}_i^\dagger \op{E}_j\ket{c_m}&=&0 \text{ for } \braket{c_l}{c_m}=0\label{eqn:kl-c2},
\end{eqnarray}
\end{subequations}
which are known to be necessary and sufficient.
If the $\op{E}_i$ are Pauli operators with maximum weight $t$, then this code is a distance $2t+1$ code.

The knowledge of exactly which qubits have been corrupted is very powerful,
and in general it allows for twice as many single qubits to be corrupted
while still allowing the data to be recovered. Therefore, given some
general error correcting code with parameters\footnote{That is, it encodes $k$ qubits using $n$ qubits and being able to correct
Pauli errors of maximum weight $t$.} $[[n,k,d=2t+1]]$, one will
only be able to correct $t$ errors at unknown locations, while
being able to correct at least $2t$ erasures\footnote{The same can be
said about classical error correcting codes.}. This follows from
the fact that there is no need
to distinguish between error operators at different locations because
the positions of the corrupted qubits are known when dealing with
erasures. With that in mind, Grassl \etal\ \cite{grass-etal:1997}
derived modified conditions for quantum erasure codes. In the 
general Knill-Laflamme conditions, each of the error operators are taken
to have weight up to $t$, so that the product $\op{E}_i^{\dagger}\op{E}_j$
has weight up to $2t$. Condition \eqref{eqn:kl-c1} therefore says
that valid states disturbed by an error operator $\op{E}_i^{\dagger}\op{E}_j$
are scaled and rotated in the same way.
Condition \eqref{eqn:kl-c2} on the
other hand guarantees that $\op{E}_i^{\dagger}\op{E}_j$ will never map orthogonal
encoded states into non-orthogonal states, so that the different basis states,
when corrupted, are still perfectly distinguishable.
Since it is known over which qubits $\op{E}_i^{\dagger}\op{E}_j$ acts non-trivially, 
this erasure operator can also be corrected. Thus one can talk about
a set of correctable erasure operators made up of all products $\{\op{E}_i^{\dagger}\op{E}_j\}$
where $\op{E}_i,\op{E}_j$ are correctable errors. If the set of erasure operators
is relabeled $\{\op{A}_i\}$ then they satisfy the {\em Modified Knill-Laflamme
Conditions} \cite{grass-etal:1997}
\begin{subequations}
\begin{eqnarray}
\bra{c_l}\op{A}_i\ket{c_l}&=&\bra{c_m}\op{A}_i\ket{c_m}\label{eqn:mkl-c1}\\
\bra{c_l}\op{A}_i\ket{c_m}&=&0 \text{ for } \braket{c_l}{c_m}=0\label{eqn:mkl-c2},
\end{eqnarray}
\end{subequations}
making it clear that $t$ error correcting codes are in fact $2t$ erasure
correcting codes. Note that the same general method is employed to
identify which underlying Pauli operator acted on the data regardless of
whether it is an erasure or a general error, as will be demonstrated later in
this chapter. For an erasure pattern, however, the qubits over which an operator may have acted non-trivially
are known, so all the $\op{A}_i$ are only required to come from the same
erasure pattern. In essence, independently for each correctable erasure pattern,
all the Pauli operators in the convex sum describing the erasure
pattern must satisfy \eqref{eqn:mkl-c1} and \eqref{eqn:mkl-c2}. 

\section{Calderbank-Shor-Steane Codes}
One of the first large classes of quantum codes to be discovered were
codes based on pairs of orthogonal classical codes. These codes are named 
{\em CSS codes} in honor of the discoverers: Calderbank, Shor 
\cite{bank-shor:1996} and Steane \cite{steane:1996}.

To understand how CSS codes are constructed, first we need to quickly review
some basic facts about classical error correcting 
codes \cite{shor:1996}. 

A classical code $C$ is a set of length $n$ vectors over $GF(2)$ (the integers 
modulo 2). Each of these vectors is called a {\em codeword}, and if this
set forms a subspace of $GF(2)^n$ then it is called a {\em linear
classical code}. From here on classical linear codes will
be referred to simply as linear codes. The Hamming weight $wt(c)$ of a
vector $v$ is the number of non-zero elements. The Hamming
distance of two vectors $x,y$ is defined by $wt(x-y)$. Recall
that, because we are in $GF(2)$, $1+1=1-1=0$, so the operation
can be thought of as an elementwise XOR.

A linear code $C$ with $2^k$ codewords of length $n$ is said to be a
$[n,k,d]$ code if the minimum weight of a non-zero codeword is $d$ --
this is also the minimum distance between two distinct codewords.
Because this is a linear subspace of dimension $k$, we can
describe the code by a generator matrix $G$ with dimensions $k\times n$, 
where the rows are linearly independent
codewords, and then every uncoded length $k$ row vector $v$ can be encoded into
a codeword $c$ by performing the matrix multiplication
\begin{equation}
vG=c.
\end{equation}
The same code $C$ may be described implicitly by the codespace
orthogonal to it, $C^\perp$. That is, we can construct a $n-k$ by $n$ matrix
$H$ such that 
\begin{equation}
Hc=0,
\end{equation}
if $c$ is any valid codeword in $C$.

The minimum distance of a linear code is important because
any vector over $GF(2)^n$ can be associated with at most one
valid codeword if it is within a Hamming distance of 
\begin{equation}
t=\left\lfloor\frac{d-1}{2}\right\rfloor, 
\end{equation}
and therefore we
call a code with distance $d$ a $t$ error correcting code.
If all we are interested in is detecting whether an error
has occurred or not, we can tolerate at most $d-1$ errors,
since $d$ errors may lead one valid codeword into a different one
without being detected.

Since it is known that $Hc=0$ for any valid codeword, if we consider
some error vector $e$, then we have
\begin{equation}
H(c+e)=Hc+He=He.
\end{equation}
As long as $e\not\in C$, this is a non-zero value, and it will tell us
that an error has occurred -- we have {\em detected} an error. If we
assume $wt(e)\le t$, then the value $He$ is uniquely mapped to $e$, 
and one may simple apply $e$ to the corrupted data to obtain
$c+e+e=c$.

The $n-k$ rows of $H$ are linearly independent vectors that are
orthogonal to all valid codewords in $C$, as previously stated. So these
vectors may be thought of as a basis for the orthogonal codespace $C^{\perp}$
(which may share more than the zero codeword with $C$).

Consider the case where $C^{\perp}\subseteq C$ are linear
codes with $n$ bit long codewords\footnote{This is 
not the most general CSS code construction, but for simplicity
we restrict ourselves to this case.}. We say that
two codewords in $C$ are equivalent if they differ by
an element of $C^{\perp}$. If $C$ is a $[n,k,d]$ code, 
then $C^{\perp}$ encodes $n-k$ bits, and by a simple
counting argument, it is clear that there must be $2k-n$
equivalence classes of $C^\perp$ in $C$. We define the 
quantum superposition $\ket{i_c}$, for some $c\in C$, to be
\begin{equation}\label{eqn:css-super}
\ket{i_c}=\frac{1}{\sqrt{\left|C^{\perp}\right|}}\sum_{u\in C^{\perp}}\ket{c+u}.
\end{equation}
This is, in effect, a superposition over the equivalence class
containing $c$. It is clear that if $c_1$ and $c_2$ are
non-equivalent classical codewords
\begin{equation}
\braket{i_{c_1}}{i_{c_2}}=0,
\end{equation}
and since there are $2k-n$ of these, we consider the space spanned by
collection of all possible $\ket{i_c}$ to
be a $[[n,2k-n]]$ quantum code. Since the codewords of $C$ are guaranteed
to have minimum distance $d$, $\ket{i_c}$ will always be
orthogonal to a state resulting from the application of less than
$d$ bitflips, or $\X$ operators.

Applying the qubitwise Hadamard transform to \eqref{eqn:css-super} one 
obtains\footnote{By using the identity
\begin{equation}
  \sum_{c\in C}(-1)^{c\cdot u}=\left\{
\begin{array}{ll}
2^k & \text{ if $u \in C^{\perp}$}\\
0 & \text{ if $u \not\in C^{\perp}$.}
\end{array}\right.
\end{equation}}
\begin{equation}\label{eqn:x-basis-css}
\ket{\tilde{i}_c}=\frac{1}{\sqrt{\left|C\right|}}\sum_{u\in C}(-1)^{c\cdot u}\ket{u},
\end{equation}
which is a superposition over codewords of $C$ with relative phases
dependent on $c$. Again, because $C$ has minimum distance $d$, if we apply
less than $d$ bitflips to $\ket{\tilde{i}_c}$ we obtain a state 
orthogonal to it, which is not the case if we apply $d$ bitflips.
Because we are in the Hadamard basis, that means that the $\ket{i_c}$
basis can withstand at most $d-1$ phase flips, or $\Z$ operators, and the
error will still be detected. By analogy, we define this to be a $[[n,2k-n,d]]$
quantum code, since there are weight $d$ Pauli operators that cannot be 
detected as errors.
 
\section{Stabilizer Codes}
Developed by Gottesman\cite{gottesman-thesis:1997}, 
and by Sloane, Shor, Calderbank and Rains\cite{calderbank-etal:1998} under
a different formalism, stabilizer codes are a class of quantum
codes much broader than the one described by CSS codes.

The basic idea is to define an Abelian subgroup $S$ of the $n$ qubit Pauli group
$\Pauli_n$, and to take the common eigenspace with eigenvalue +1
of $S$ as the code space. $S$ is called the stabilizer of the code,
since any operator $\op{M}\in S$ and encoded state $\ket{\enc{\psi}}$
gives
\begin{equation}
\op{M}\ket{\enc{\psi}}=\ket{\enc{\psi}},
\end{equation}
that is, all elements of $S$ act trivially on the codespace.
It can be shown that if $S$ has $n-k$ generators, then the
codespace has dimension $2^k$, and therefore it encodes
$k$ qubits.

Now consider an error operator $\op{E}\in\Pauli_n$. If for some $\op{M}\in S$
we have\footnote{Recall that $\{\op{A},\op{B}\}$, the anti-commutator
between $\op{A}$ and $\op{B}$, is defined as $\{\op{A},\op{B}\}=\op{AB}+\op{BA}$.} $\{\op{M},\op{E}\}=0$, then
\begin{equation}\label{eqn:error-eigenvalue}
\op{ME}\ket{\enc{\psi}}=-\op{EM}\ket{\enc{\psi}}=-\op{E}\ket{\enc{\psi}}
\end{equation}
so $\op{E}\ket{\enc{\psi}}$ is a $-1$ eigenstate of $\op{M}$, and this can be
detected by measuring $\op{M}$. Going back to the modified Knill-Laflamme
conditions for erasure correction, one finds that if $\op{A}_i\in S$,
then \eqref{eqn:mkl-c1} holds, since 
\begin{equation}
\bra{c_l}\op{A}_i\ket{c_l}=\braket{c_l}{c_l}=1,
\end{equation}
for any encoded states $\ket{c_l}$. If $\{\op{A}_i,\op{M}\}=0$  for
some $\op{M} \in S$ then \eqref{eqn:mkl-c2} holds, since
\begin{equation}
\bra{c_l}\op{A}_i\ket{c_m}=\bra{c_l}\op{A}_i\op{M}\ket{c_m}=-\bra{c_l}\op{M}\op{A}_i\ket{c_m}=0.
\end{equation}
so we can restrict ourselves to looking at the commutation relations of
erasure operators to determine their correctability.

\subsection{CSS Codes as Stabilizer Codes\label{subsec:css-as-stab}}
From the basis state representation of CSS codes, we can infer what the
stabilizer of the code should be. Recall that the equivalence relation
that defines the $\ket{i_c}$ in \eqref{eqn:css-super} is that if two 
codewords in $C$ differ by an element of $C^{\perp}$, then they are equivalent.
Clearly, if we add some $w\in C^{\perp}$ to each of the classical codewords in
$\ket{i_c}$ the state remains unchanged -- all codewords remain in the
same equivalence class. We can take this operation to be a bit flip operator $\op{W}_\X$,
obtained by replacing each $0$ in $w$ by an $\I$, and each $1$ is replaced
by an $\X$, all elements concatenated by tensor products. This is
a set of bit flip operators in the stabilizer, and they can be generated
by the $n-k$ linearly independent bitflip operators obtained from
the $n-k$ generators of $C^{\perp}$.

Looking at the Hadamard basis description of the CSS codes as given
in \eqref{eqn:x-basis-css}, we notice
a similar fact, with a little more algebra involved. Again, if we
apply a bit flip operator $\op{W}_\X$ based on a codeword $w\in C^{\perp}$,
we obtain
\begin{subequations}
\begin{eqnarray}
\op{W}_{\X}\ket{\tilde{i}_c}&=&\frac{\op{W}_{\X}}{\sqrt{\left|C\right|}}\sum_{u\in C}(-1)^{c\cdot u}\ket{u}\\
&=&\frac{1}{\sqrt{\left|C\right|}}\sum_{u\in C}(-1)^{c\cdot u}\ket{u+w}\nonumber\\
&=&\frac{1}{\sqrt{\left|C\right|}}\sum_{u'\in C}(-1)^{c\cdot (u'-w)}\ket{u'}\nonumber\\
&=&\frac{1}{\sqrt{\left|C\right|}}\sum_{u'\in C}(-1)^{c\cdot u'-c\cdot w}\ket{u'}\nonumber\\
&=&\frac{1}{\sqrt{\left|C\right|}}\sum_{u'\in C}(-1)^{c\cdot u'}\ket{u'}\nonumber\\
\op{W}_{\X}\ket{\tilde{i}_c}&=&\ket{\tilde{i}_c}
\end{eqnarray}
\end{subequations}
Thus $\op{W}_\X$ stabilizes the $\ket{\tilde{i}_c}$ as well. In the
computational basis, this operator is the same, except that the $\X$s
are replaced by $\Z$s, so we write it as the operator $\op{W}_\Z$. Again, there
are $n-k$ linearly independent operators such as these, since there are
$n-k$ generator codewords for $C^{\perp}$.

Taking all the possible $\op{W}_\Z$ and $\op{W}_\X$, we have $2n-2k$
linearly independent generators, and we have $n$ qubits. Thus, with
these generators we encode $n-(2n-2k)=2k-n$ qubits, which is
exactly the number of qubits that the CSS code encodes, so the
$\op{W}_\Z$ and $\op{W}_\X$ generate {\em all stabilizer operators} of the
CSS code.

For any self-orthogonal linear code $C$ with parity check matrix $H$,
we obtain a CSS code with stabilizer generators obtained from $H$
as described in such a way that we have $n-k$ generators that are
tensor products of only $\X$s and $\I$s (what we call the
$\X$ stabilizers), and $n-k$ generators that are
tensor products of $\Z$s and $\I$s (what we call the $\Z$ stabilizers).
\subsection{The Normalizer and the Heisenberg Representation\label{subsec:norm-and-heisenberg}}
There are Pauli operators that commute with all elements of the 
stabilizer $S$ but that do not necessarily leave the codespace invariant. 
That is,
there are operators $\op{O}\in \Pauli_n$ such that
\begin{subequations}
\begin{eqnarray}
\forall \op{M}\in S, \text{  }[\op{M},\op{O}]&=&0\\
\op{O}\ket{\enc{\psi}}&=&\alpha\ket{\enc{\chi}},
\end{eqnarray}
\end{subequations}
where $\ket{\enc{\psi}},\ket{\enc{\chi}}$ are encoded basis states (possibly equal)
and $\alpha=\pm 1$. 
The set of such operators is called the {\em normalizer of} $S$, denoted
$N(S)$. Operators in $N(S)/S$ are errors that cannot be detected because
they map valid codewords into different valid codewords, and
because of that they can also be seen as encoded operations on the
encoded data -- in fact, they are the encoded Pauli operations acting on 
encoded qubits.

Observing the evolution of $S$ and $N(S)$ under the action of different
unitary operators can be used to determine the behaviors of
certain types of circuits, and it is especially helpful in
constructing encoded operations \cite{gottesman:1998,zlc:2000}.
This is what is called the {\em Heisenberg representation of quantum
computers} \cite{gottesman-heisenberg:1998}, since it is based on the general
idea of tracking the evolution of the operators in $N(S)$ -- $S$ simply
being a particular subset of $N(S)$ -- much like
one tracks the evolution of operators in the Heisenberg picture of
quantum mechanics. The general idea is to observe how some operators
$\op{M}_i\in N(S)$ evolves under the action of some unitary operator $\op{U}$, by noting
in particular that
\begin{subequations}
\begin{eqnarray}
\op{U}\op{M}_i\ket{\psi}&=&\op{U}\op{M}_i\op{U}^\dagger \op{U}\ket{\psi}\label{eqn:norm-1}\\
\op{U}\op{M}_i\op{M}_j\op{U}^\dagger&=&\op{U}\op{M}_i\op{U}^\dagger \op{U}\op{M}_j\op{U}^\dagger\label{eqn:norm-2}.
\end{eqnarray}
\end{subequations}
We may interpret \eqref{eqn:norm-1} as a statement of what $\op{M}_i$ is
mapped to under the action of $\op{U}$, so that, for example, we may know
how its eigenstates get mapped under the action of $\op{U}$. On
the other hand, \eqref{eqn:norm-2} allows us to restrict our attention
to a generating set of unitary operations -- the action of the generated group
follows by linearity.

However, in general, a unitary
operator will not map a Pauli operator into another Pauli operator.
We can consider, however, a particular class of unitary operators
that is very useful.
\begin{definition}
The {\bf Clifford  Group}, denoted $\C_2$, is the set of unitary 
operators that maps the Pauli group into itself under conjugation 
\cite{gottesman-chuang:1999}. That is
\begin{equation}
\C_2 = \{\op{U} | \op{UOU}^{\dagger}\in\Pauli_n \text{ for all } \op{O} \in \Pauli_n\}.
\end{equation}
\end{definition}
Since $S\subset\Pauli_n$ and $N(S)\subset\Pauli_n$, we can consider
circuits made up of gates in $\C_2$, and the Heisenberg representation
allows us to monitor the evolution of the encoded states by observing
the evolution of operators that generate the encoded Pauli set
$\Pauli_k$.

$\C_2$ is finitely generated by the Pauli group
plus the Hadamard, $\Had$, the phase gate $\Pha=\text{diag}(1,i)$, 
and the $\CNOT$.
If we 
restrict our attention to the action of these gates, we easily find 
how any normalizer is transformed. For the purposes of this
thesis, it suffices to look at the action of the generating set of
unitary gates of $\C_2$.

The Hadamard gate maps the Pauli operators as
\begin{subequations}
\begin{eqnarray}
\Had\X\Had &=& \Z\\
\Had\Z\Had &=& \X,
\end{eqnarray}
\end{subequations}
so that, for example, we may consider an $\X$ followed by a $\Had$
to be the same as a $\Had$ followed by a $\Z$. This, in fact, is
a very useful tool in observing the propagation of error operators
in quantum circuits. The mappings for the other gates in the
generating set of $\C_2$ are
\begin{subequations}
\begin{eqnarray}
\Pha\X\Pha^\dagger&=&i\Y\\
\Pha\Z\Pha^\dagger&=&\Z\\
\CNOT_{1,2}(\X\ox\I)\CNOT_{1,2}&=&\X\ox\X\\
\CNOT_{1,2}(\I\ox\X)\CNOT_{1,2}&=&\I\ox\X\\
\CNOT_{1,2}(\Z\ox\I)\CNOT_{1,2}&=&\Z\ox\I\\
\CNOT_{1,2}(\I\ox\Z)\CNOT_{1,2}&=&\Z\ox\Z,
\end{eqnarray}
\end{subequations}
along with the identity
\begin{equation}
(\I\ox\Had)\CNOT_{1,2}(\I\ox\Had)=\CSIGN_{1,2},
\end{equation}
will be used extensively in the construction of fault-tolerant
encoded gates in Chapter \ref{ch:candidates}. Note that if 
we consider instead rotations about Pauli operators,
then all $90^\degg$ rotations and its integral powers will also be in
$\C_2$. These, however, are just a different representation of the gates
that can be generated by the set of
gates described above.
\section{Erasure Correction Procedure\label{sec:era-corr}}
The process of erasure correction
hinges on the measurement of the stabilizer operators for the code,
like in error correction codes, 
but now there is the extra knowledge of where the erasure occurred.
Here we follow an approach proposed by Zalka \cite{zalka:personal} and
based on the work of Shor \cite{shor:1996}. The general idea is that,
given a single erasure in an erasure pattern which may include several, one
attempts to measure a stabilizer that acts trivially on all other
erasures, but that acts non-trivially on the erasure that is being
targeted for correction -- what is called the {\em correction target} --
and on qubits unaffected by erasures. The outcome of the measurement,
which is described in more detail in the next section, indicates
which action must be taken to correct the erasure.

In the case of a $\Z$ erasure, it is necessary to determine whether a
$\Z$ error has indeed been applied, or if the identity has been
applied. Thus, it is sufficient to measure a stabilizer operator
with an $\X$ on the same position as the correction target. In the
case of a full-erasure, the erasure needs to be corrected in two
separate steps. The reason for that is clear by noting that
\begin{subequations}
\begin{eqnarray}
\er{X}(\er{Z}(\rho))&=&\frac{1}{2}\left[\er{Z}(\rho)+\X\er{Z}(\rho)\X\right]\\
&=&\frac{1}{4}(\rho+\X\rho\X+\Z\rho\Z+\Y\rho\Y)\\
&=&\er{E}(\rho)\\
&=&\er{Z}(\er{X}(\rho)),
\end{eqnarray}
\end{subequations}
so we may consider a full erasure to be an $\X$ erasure followed by
a $\Z$ erasure. First we measure a stabilizer with a
$\Z$ on the correction target position, and correct for an $\X$ erasure,
and then we measure a stabilizer with a $\X$ on the correction target position,
correcting for a $\Z$ erasure in that position.

When considering a single erasure, there is no difference between choosing to
correct the $\Z$ erasure or the $\X$ erasure first. However, when considering
an erasure pattern that consists of multiple erasure of multiple type, the
best strategy is to make a choice that is least likely to lead to an uncorrectable
error. This depends both on the error model and on the code being used,
and will be discussed in more detail in Chapter \ref{ch:candidates}, after both
topics have been introduced.
\section{Fault-tolerant Computation}
We would like to be able to perform useful computation
on a quantum computer regardless of how long the computation is or how many
qubits are involved, simply because
we would like to solve many different types of problems, of different
complexities, with different input sizes.
If one expects the
error rate of the quantum computer to be naturally low enough
so that errors are unlikely to occur during computation, one
finds that the acceptable error rates are dependent on the
size of the computation. 
Thus, we'd like a means to perform any useful quantum computation 
even in the presence of a fixed probability of error for each
gate.
This is what is generally meant
by {\em fault-tolerant quantum computation}.

Encoding the data to resist error operators is not enough to
reach this objective. If, in
order to do computation, one was required to decode the data, perform
computation, and then re-encode the data, there would be no protection
from the  noise and decoherence during the computational
step. Universal computation with the protection of error correction
codes is not trivial, however, since it requires that the encoded
operations necessary for universal computation be identified. It is
not even sufficient to perform these encoded operations correctly,
because it is possible that the computation still allows errors to
propagate in a catastrophic way -- it could be that even during a step
where no errors have  occurred, careless computation could take a
correctable error into an uncorrectable error. The canonical example
of this is the $\CNOT$ gate. It has been stated before  that $\CNOT$
is in the Clifford group, and the exact mapping it performs -- as
seen in the previous section -- does not preserve the weight of
length two Pauli operators.
Clearly, if we perform the $\CNOT$ between two qubits of the same code
block, we run the risk of increasing the weight of the error, possibly
leading to an uncorrectable error, even when none of the $\CNOT$s fail,
simply because the data contained errors that were propagated
carelessly. A general rule that can be extracted from this is that
we do not want errors to propagate within a code block, so we do not
allow for qubits in the same code block to interact with each other.
This, essentially, translates to the requirement that encoded gate
operations
be transversal -- that they operate qubitwise on a code block.

\subsection{Fault-tolerant Stabilizer Measurement\label{subsec:ft-stab-meas}}
\begin{figure}
\centering
\includegraphics{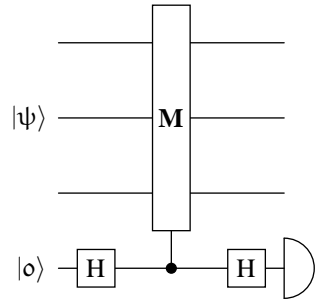}
\caption{Quantum circuit for measuring the eigenvalue of a stabilizer 
operator.\label{fig:phase-kickback}}
\end{figure}

In stabilizer codes, in order to determine which error has affected the data,
one needs to measure some subset of stabilizer operators, and in general the
stabilizer generators suffice. According to \eqref{eqn:error-eigenvalue},
we know that a detectable error has eigenvalue $-1$. On the other hand, the absence of
errors or undetectable errors have an eigenvalue $+1$, so we can use
the phase kick-back quantum circuits to measure the eigenvalue
of the data, as depicted in Figure \ref{fig:phase-kickback}. The control qubit
will be in the state $\ket{\logic{0}}+\ket{\logic{1}}$ after the Hadamard gate.
Because $\op{M}$ is a stabilizer of the codespace, whether it is applied
or not does not affect valid data at all, and if there is no error (or if the
error commutes with $\op{M}$), both
control and data are unaffected, the second Hadamard brings the control back to
$\ket{\logic{0}}$ and that is the state that is measured.
If  there is an error that anticommutes with $\op{M}$, then 
the control will be 
in the state $\ket{\logic{0}}-\ket{\logic{1}}$, and the second Hadamard will
bring it to $\ket{\logic{1}}$ which is then detected. Thus, clearly
detecting a $\ket{\logic{0}}$ indicates commutation, while a $\ket{\logic{1}}$
indicates anticommutation.

\begin{figure}
\centering
\includegraphics[scale=0.7]{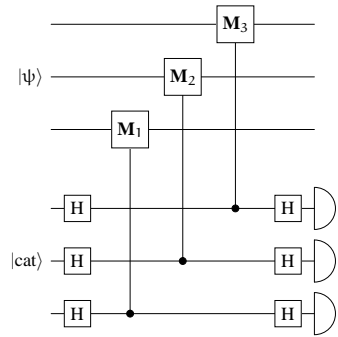}
\caption[Quantum circuit for fault-tolerant stabilizer measurement.]{Quantum circuit for measuring the eigenvalue of a weight three stabilizer operator 
$\op{M}=\op{M}_1\ox\op{M}_2\ox\op{M}_3$
fault-tolerantly. This generalizes in the obvious way for stabilizers of
higher weight.\label{fig:ft-phase-kickback}}
\end{figure}

Because the data is made up of multiple qubits, and the stabilizer acts non-trivially
over more than one of them, the circuit in Figure \ref{fig:phase-kickback} is not fault-tolerant.
This is because if one of the $\CNOT$s or $\CSIGN$s (for $\X$s and $\Z$s in $\op{M}$
respectively, not shown in the figure) fails, 
it can cause an error on the control bit, which is shared by
all the other controlled operations on other qubits. For example, if a failure
occurs on the first controlled operation, {\em all} other qubits over which
$\op{M}$ acts non-trivially will be affected.

The solution for this problem is quite simple. If we replace the single control qubit
by a group of qubits, one for each controlled operation needed to measure the
stabilizer $\op{M}$, and interact with the data transversally as depicted in
Figure \ref{fig:ft-phase-kickback}, there is
no catastrophic error propagation. In this case, the control lines need to be
replaced by a 'cat' state, of the form
\begin{equation}
\ket{\logic{0}}^{\ox m}+\ket{\logic{1}}^{\ox m},
\end{equation}
where $m$ is the number of qubits over which $\op{M}$ acts non-trivially. If the
error anticommutes with $\op{M}$, this cat state will develop a relative phase
of $-1$. 
\begin{equation}\label{eqn:bad-cat}
\ket{\logic{0}}^{\ox m}-\ket{\logic{1}}^{\ox m}.
\end{equation}
Recall, however, equation \eqref{eqn:x-basis-css}, and take $C$ to be the
$m$ bit repetition code -- that is, a code that maps 
\begin{subequations}
\begin{eqnarray}
0&\to&\underbrace{00\cdots 0}_{\text{m times}}\\
1&\to&\underbrace{11\cdots 1}_{\text{m times}}. 
\end{eqnarray}
\end{subequations}
If we apply the qubitwise Hadamard to \eqref{eqn:bad-cat},
we obtain a superposition of odd weight binary strings, and by measuring
and computing the parity classically, we can fault-tolerantly detect that 
an error has occurred.

The use of cat states is necessary because it ensures that no information
about the state of valid data is transfered to the ancilla. This is important
because by measuring the ancilla we do not want to collapse the superposition
of valid encoded states with no errors.
\section{Threshold Theorem}
Once the data is protected by an erasure correction code, the probability
of erasure on the encoded data may still be unacceptably high. One way to get
around this problem is to perform {\em concatenated coding}, that is,
encode the encoded data once again. Say, for example, we have a code
$C^{(1)}$ with parameters $[[n,1,d]]$. By concatenating the code with itself once,
we obtain a $[[n^2,1,d^2]]$ code which we call $C^{(2)}$, on which erasure correction can be seen
as erasure correction on each level of concatenation separately. It is
straightforward to see why the minimum distance scales in such a way. If $C$ has as its encoded
operators $\enc{\X}$ and $\enc{\Z}$, then by replacing the uncoded
Pauli operators by these encoded ones in the stabilizer of $C$, we obtain a 
new code $C^{(2)}$, which also includes the $n$-fold Cartesian product 
of the stabilizer of $C$. This procedure can be repeated $L$ times to obtain the concatenated code
$C^{(L)}$ with parameters $[[n^L,1,d^L]]$.

Consider $C^{(1)}$ once again. If the qubit erasure rate is $\e$, then we can write the
block failure rate as
\begin{equation}
\e^{(1)}=\sum_{i=d}^{N}c_i\e^i\label{eqn:rec},
\end{equation}
where $c_i$ are integer coefficients dependent only on the erasure correction procedure
and $i$. We are only interested in the case $d>2$, since we want to be able to correct
at least one erasure. This recursion relation can be calculated by analyzing how erasures
are introduced during the erasure correction procedure, and how this may lead to
an uncorrectable erasure pattern in a code block, what we call a {\em failure}. 
In theory, one often assumes that erasure correction is 
attempted until the data is erasure free, leading $N$ to be infinite. In practice,
however, only a certain maximal number of erasure correction steps are attempted,
placing a bound on $N$.

Taking only the first term of the recursion relation \eqref{eqn:rec}, we can
approximate the concatenated block failure rate as
\begin{equation}
\e^{(L)}\approx c_d^L\e^{d^L}.\label{eqn:scale}
\end{equation}
In the case that $\e^{(1)}<\e$, then \eqref{eqn:scale} indicates that
the error rate of a concatenated code will drop doubly exponentially
with $L$, while the size of the code block grown exponentially. This is,
in essence, what is called the {\em threshold theorem} 
\cite{aharonov,klw,preskill}, which holds under various different conditions,
but for our purposes it suffices to say that there is an limitless supply
of fresh qubits, that the base erasure rate is independent of the size of
the circuit, and that the probability of erasures on the different gates
are independent. The value of $\e$ which gives $\lim_{L\to\infty}\e^{(L)}=0$ is
called the erasure threshold, and obtaining such a value is the main
focus of this thesis. In practice, this can be taken as the value of
$\e$ which gives $\e^{(1)}<\e$, as long as we assume that gate failures are
independent, that the gate failure rate does not depend on the computation size,
and that fresh qubits can be produced on demand. A much more detailed description of how to
obtain the recursion relation \eqref{eqn:rec} and how to extract
the threshold will be given in Chapter \ref{ch:erasure-threshold}.

%% file: eLOQC.tex
\chapter{Efficient Linear Optics Quantum Computation\label{ch:eloqc}}
\markright{Efficient Linear Optics Quantum Computation}
A very brief overview of the efficient linear optics quantum
computing proposal by Knill, Laflamme and Milburn \cite{klm:2001}
is given, focusing on the behavior instead of implementation
details. From that description, an error model is derived.
\section{The Knill-Laflamme-Milburn proposal}
One of the earliest proposals for a quantum computer, put forth
by Chuang and Yamamoto \cite{chuang-yama:1995}, described
a system where qubits were encoded in two photon modes -- the so called
{\em dual rail encoding}. In order to differentiate quantum states
representing photon number states and quantum states representing the
qubit, we follow the convention of using the standard font for
number states, and the gothic font for qubits, so that the
dual rail encoding of the $i$th qubit would be represented by
\begin{subequations}
\begin{eqnarray}
\ket{\logic{0}}_i & = &\ket{0}_{a_i}\ket{1}_{b_i}\\
\ket{\logic{1}}_i & = &\ket{1}_{a_i}\ket{0}_{b_i},
\end{eqnarray}
\end{subequations}
where $a_i$ and $b_i$ are the modes corresponding to qubit $i$.
The main motivation of using photons to encode qubits is that photons
do not decohere as easily as most other physical systems used
to implement qubits, simply because photons can easily be made to not 
interact strongly with the environment.

One can use very simple optical elements to perform single qubit 
operations, and this was constructively proven well before the
Chuang-Yamamoto proposal \cite{zeilinger:1994}. The elements used
are called {\em passive linear optics elements}, and they are
comprised of beam splitters (partially reflective mirrors) 
and phase shifters (delays). These elements have the property that
they preserve the number of photons, and their behavior can easily
be described by how they transform the photon creation operators
of the modes involved.

However, it is very hard to make a universal
set of quantum gates. In any universal set there is at least one
entangling gate that requires the interaction of different qubits, and
if the qubits are encoded as photons, it is very hard to construct such 
gates without using non-linear media to mediate the interaction. In
\cite{chuang-yama:1995}, a Kerr non-linear medium was proposed
to construct entangling gates, but Kerr media are notorious for
having very high loss. Even measurements that require an implicit
interaction of the qubits, such as Bell-basis measurements, are
impossible to perform without failure \cite{lutkenhaus:1999}.
It was thought that these facts comprised an informal
``no-go'' theorem for linear optics quantum computing, and various
proposals that attempted to build quantum computers out of
linear optics were shown to require an exponential amount
of physical resources.

When Gottesman and Chuang \cite{gottesman-chuang:1999} demonstrated
that fault-tolerant universal quantum computation could be performed by
using quantum teleportation and state preparation, this picture changed.
The state preparation required for this gate construction scheme
depends only on the gate being implemented, not on the inputs to the gate,
so that if there is a way to prepare these states offline, linear
optics quantum computation can indeed be performed (a fact that was
pointed out in \cite{gottesman-chuang:1999}). One just needs to keep
trying to prepare this states until the preparation succeeds, or
maintain many copies of the successfully prepared states.

Building on these ideas, Knill, Laflamme
and Milburn put forth what is now commonly referred to as the KLM
proposal for efficient linear optics quantum computation \cite{klm:2001}.
This proposal uses single photon sources and detectors, linear
optics, and post-selection based on measurement outcomes.

There are two crucial parts of this proposal: the state preparation,
and efficient teleportation. It was shown that one can build a
non-deterministic non-linear sign change operation on a single photon mode,
with knowledge of success or failure,
using only linear optics and measurements. This gate, called
$\NS_{-1}$, in effect performs the transformation
\begin{equation}
\ket{0} + \ket{1} + \ket{2} \to \ket{0} + \ket{1} - \ket{2},
\end{equation}
with a finite probability of success, which is detected
by measuring some ancillary modes in the gate construction. Loose bounds
have been placed on the success probability of constructions of
this gate using only linear optics \cite{knill-bounds:2003},
and it is known that these gates cannot succeed with a probability
of $\frac{1}{2}$ or more. One can construct a
controlled phase flip, also known as a controlled sign or $\CSIGN$,
using applications of the $\NS_{-1}$ and linear optics elements 
on the dual rail encoding. 
Direct computation with these gates is
not scalable, but using the teleportation scheme proposed
by \cite{gottesman-chuang:1999}, scalability is achieved by using
these gates {\em only} in state preparation, where numerous attempts
can be made until the gate performs the desired operation.

There is still the problem of how teleportation is performed. In the
standard teleportation protocol, one needs entangled qubit
pairs, which can be prepared offline, as well as Bell measurements. We have 
mentioned already that \cite{lutkenhaus:1999} showed that such measurements
are impossible to perform without probabilistic failure -- in
fact, one cannot distinguish between two of the states in the four
state Bell basis, and because of that, the probability of failure
cannot be made less than $50\%$ if the input states are chosen
uniformly over the Hilbert space. The alternative proposed in
\cite{klm:2001} is to use a modified protocol that relies
on the preparation of a larger ancillary entangled state, 
the application of a
Fourier transform involving the data to be teleported and
the prepared state, and measurement of
the ancillary states -- the Fourier transform can be implemented
efficiently using only linear optics. Given an ancillary state consisting of
$N$ qubits, this teleportation protocol succeeds with probability
$\frac{N}{N+1}$, a great improvement over the standard teleportation
protocol.
Note that $N=1$ is the base case with $\frac{1}{2}$ probability of
success. This means one qubit is used for the teleportation, which seems
to disagree with the knowledge that a Bell state is necessary for teleportation.
However, only two of the four modes needed for the two qubit $\CSIGN$
operation interact with the $\NS_{-1}$ gates, so those are the only modes that
need to be teleported, and each one requires two modes in a Bell state,
resulting in one qubit per mode teleported.

\section{The Error Model\label{sec:error-model}}
Errors are introduced into computation in \eLOQC\ through various sources:
failure of $\CSIGN$, photon loss, and finite accuracy of 
phase-shifters and beam-splitters, {\em etc}.
Here we consider only failures during the application of a $\CSIGN$ that
can be detected by measurement of the ancillary modes. In effect,
this is not a detection of the failures of the $\NS_{-1}$ gates, but
instead a detection of failures that occur during the teleportation
of the qubits that realizes the $\CSIGN$ given the appropriate
ancillary state. The $\NS_{-1}$ gates are used only in the state preparation
for the teleportation, so we can choose to simply not use states
resulting from failures of these gates.

Two failure modes are discussed here: teleportation
failures that are inherent in the protocol because of limitations
in linear optics, and failures due to photon loss at the detectors
during teleportation.
\subsection{Error model for ideal hardware\label{sec:ideal-error-model}}
Assuming that all linear optics elements, all detectors and all
sources are perfect, failures are still possible in the KLM
proposal. This is due to the fact that teleportation succeeds with a
probability less than unity, although these failures are always detected.

\begin{figure}
\centering
\subfigure[Teleportation followed by a $\CSIGN$.]{
\includegraphics[scale=0.7]{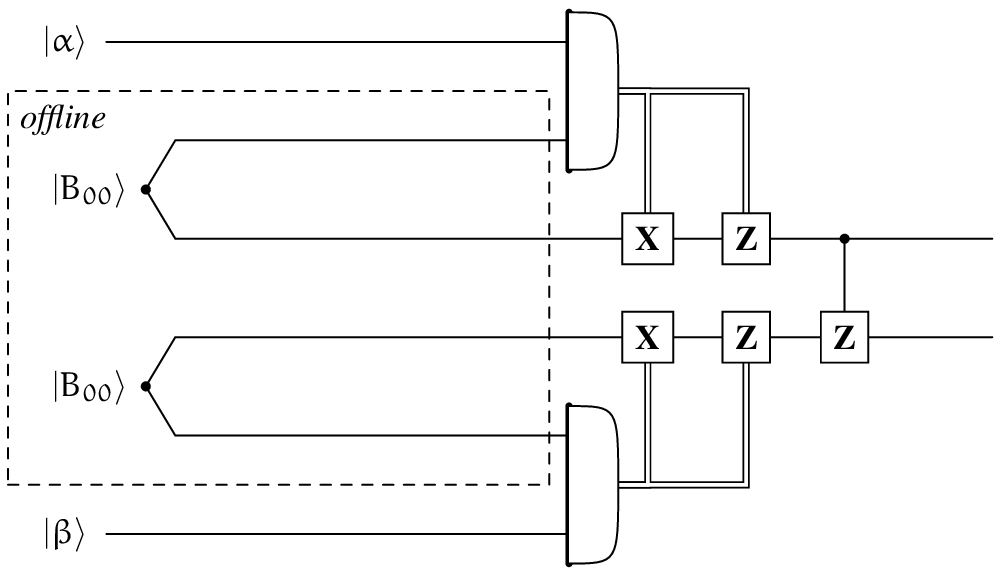}
}
\subfigure[$\CSIGN$ commuted backward into the prepared state,
which is equivalent to (a).]{
\includegraphics[scale=0.7]{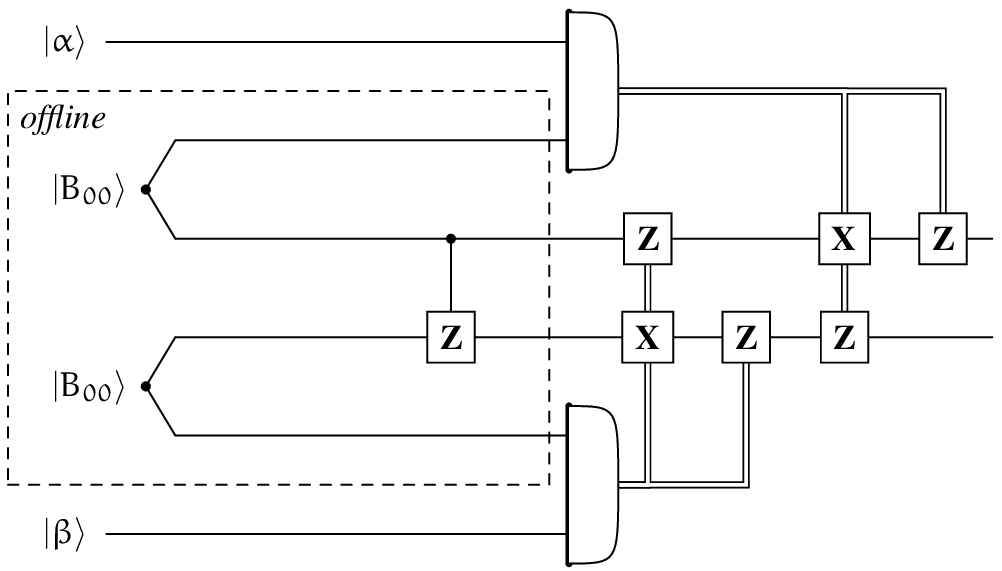}
}
\caption[The teleportation of a $\CSIGN$ gate.]{The teleportation of a $\CSIGN$ gate \protect\cite{gottesman-chuang:1999,klm:2001}.
The states $\ket{B_{00}}$ are, in the simplest case,
the Bell pair $\frac{1}{\sqrt{2}}\left(\ket{\logic{00}}+\ket{\logic{11}}\right)$. 
The two qubit measurement
is performed over the Bell basis. Different Pauli corrections are applied
to the output bit depending on the measurement outcomes, and this is indicated
by the wider lines connected to the Pauli gates.\label{fig:csign-teleport}}
\end{figure}

Consider how the $\CSIGN$ is teleported, abstracting from the
details of linear optics, as depicted in Figure \ref{fig:csign-teleport}.
The basic idea is to consider the teleportation of two qubits, followed
by the application of a $\CSIGN$ between them. The commutation relations
of $\CSIGN$ and the gates used for teleportation are used to
rewrite the circuit in such a way that the $\CSIGN$ is applied
between the prepared states used for the teleportation -- see
Section \ref{subsec:norm-and-heisenberg}.

In general, the corrections dependent on the measurement outcomes must be
applied to both qubits being teleported. However, it has been shown
that the state preparation can be modified so that, if the teleportation
of either qubit fails, it affects only that qubit. The protocol can
be further engineered in a manner such that the failure can be
taken to have occurred {\em after} the teleportation \cite{klm:2001}.
The effect of this type of failure, which is always detected, is
that of a $\Z$ measurement of the qubit that is being teleported,
with the measurement outcome made evident through the measurement part of
the teleportation protocol.
This occurs with probability $\frac{1}{N+1}$, where $N$ is the size
of the ancilla state, which in the case of $N=2$ is simply the
Bell state $\frac{1}{\sqrt{2}}\left(\ket{\logic{00}}+\ket{\logic{11}}\right)$.

One could, in principle, simply ignore the measurement outcome
and take the teleportation failure to be a $\Z$ erasure, because
the distinction between the two is classical information and
classical processing, which we take to be perfect. Formally,
this can be shown by the following proposition.

\begin{proposition}
A projective $\Z$ measurement of unknown outcome at a known
location is equivalent to a phase erasure.\label{prop:z-erasure}
\end{proposition}
\begin{proof}
Expand the probabilistic application of the two possible projection operators
for the eigenvectors of $\Z$
\begin{subequations}
\begin{eqnarray}
\Z_{+} &=& \frac{1}{2}(\I+\Z)\\
\Z_{-} &=& \frac{1}{2}(\I-\Z),
\end{eqnarray}
\end{subequations}
onto a density matrix $\rho$. The
probability of the outcome being the $+1$ eigenstate of $\Z$ is denoted
by $\Pr(\Z_+)$, and for the $-1$ eigenstate is $\Pr(\Z_-)$, and since
$\Z_++\Z_-=\I$, $\Pr(\Z_+)+\Pr(\Z_-)=1$. Ignoring the outcome
of the measurement, $\rho$ will be transformed into $\er{Z}(\rho)$ described by
\begin{subequations}
\begin{eqnarray}
\er{Z}(\rho)= &\Z_{+}\rho\Z_{+}+\Z_{-}\rho\Z_{-} \\
= &\frac{1}{4}(\I+\Z)\rho(\I+\Z)+\frac{1}{4}(\I-\Z)\rho(\I-\Z)\\
= &\frac{1}{2}\left(\rho+\Z\rho\Z\right)\label{eqn:z-meas-outcome},
\end{eqnarray}
\end{subequations}
which is a $\Z$ erasure, as claimed.
\end{proof}

In this case,
one would need to take the approach outlined in Section \ref{subsec:ft-stab-meas},
and in the case of a CSS code, one would need to measure an $\X$ stabilizer in order
to determine the syndrome of this erasure. This requires the application of
$\CNOT$s, and since they are constructed from $\CSIGN$s conjugated by
Hadamards (see Section \ref{subsec:norm-and-heisenberg}), it is clear that the type of erasure
introduced would be an $\X$ erasure, not a $\Z$ erasure. This is of particular
importance for codes that can {\em only} correct $\Z$ erasures, like the 
codes used to obtain the $0.5$ threshold \cite{manny:2001,klm-thr:2000},
because it would lead to an immediate uncorrectable failure.
In our case it desirable to avoid introducing
different types of erasures, in order to simplify the analysis in Chapter \ref{ch:erasure-threshold}. 

In order not to introduce another type of erasure unnecessarily,
we cannot use the standard fault-tolerant stabilizer measurement from the
previous chapter. Instead, we extend a technique used for correcting $\Z$ measurements,
which is described in detail in the next chapter.

For the purposes of describing the error model during computation, it
suffices to say that each $\CSIGN$ has a probability $\e_i$ of performing
an unintentional $\Z$ measurement at either the control or the
target qubit, independently\footnote{The subscript $i$ standing for ``ideal''.}. For simulation purposes, we can 
model this as a $\Z$ erasure at either target or control qubits
independently, since from a $\Z$ erasure, with perfect detectors,
one can easily obtain the measurement value without incurring
any cost. Thus, with the first qubit being the control and
the second being the target, the error model will be taken as 
\begin{subequations}
\begin{eqnarray}
\text{control failure}  &:& \er{Z}\ox\I\\
\text{target failure} &:& \I\ox\er{Z},
\end{eqnarray}
\end{subequations}
were the failures are taken to occur independently with
probability $\e_i$. Details of how the measurement outcome can
be used to aid the correction step 
will be given in Section \ref{subsec:z-measurement}.

\subsection{Error model for lossy detectors}
Still assuming infinite precision in the parameters of the phase-shifters
and beam-splitters, we can consider the possibility of photon loss due to
detector inefficiencies.
The dual-rail encoding of qubits ensures that as long as a qubit is 
properly encoded,
there is in total a single photon between the two modes
corresponding to the qubit, and since linear optics preserves
the total photon number, measurement of ancillae allows for
the detection of leakage from the dual-rail encoding.

In the original proposal for \eLOQC, a robust teleportation protocol
called ${\bf RT}_1$ is described. This protocol has the property that
it is able to detect the usual teleportation failures as well as
photon losses both in the qubits used for teleportation (ancilla  or
data) as well as in the detectors.  Like any possible linear optics
implementation of teleportation, this teleportation circuit has only a
finite probability of  success even in the ideal case, and the ancilla
measurement outcomes will indicate this type of failure that is not a
consequence of photon
loss. In the case of photon loss, however, the measurement outcomes
will be different from the case of ideal failure, and as a side
effect the modes of the qubit at the output of the teleportation are
replaced with a fresh dual-rail encoded qubit in a fixed state
\cite{klm:2001}.  This replacement by a fresh qubit corresponds to
a total loss of information about the state of the qubit over which
teleportation failed. We know that in the case of failure without
photon loss we can model the output by a phase erasure, as
demonstrated previously, but in the case of failure due to photon
loss, we have the ingredients of a full erasure: knowledge of where
the failure occurred (through the outcome of the ancilla
measurements), and complete loss of knowledge about the state of the
qubit.

We have already seen that in the case of a phase erasure, the qubit
teleportations realizing the $\CSIGN$ are affected independently -- 
if the teleportation
of the control qubit fails, it does not imply that the target qubit
teleportation fail, and vice-versa. This is not the case for the full
erasure type of failure.  Consider the teleportation of two quantum
states, and the subsequent application of a $\CSIGN$ as in
Figure \ref{fig:csign-teleport}. 
Taking a worst case approach, photon loss in the control part
of the teleported $\CSIGN$ (top half of the figure) induces a $\Z$
error at the bottom whenever there is an $\X$ error at the top as 
well (this includes $\Y$ errors, since it is a combination of
$\Z$ and $\X$ errors). A similar effect is observed in a photon
loss at target part of the teleported $\CSIGN$. Clearly there is
a classical correlation between the types of errors at the top
qubit and the types of error at the bottom qubit -- however, photon loss
occurs independently at the top and bottom part of the teleported $\CSIGN$. 
Assuming perfect and instantaneous classical processing and communication, we could exploit
this correlation in order to reduce the number of syndrome measurements
that need to be performed. Ideally, because $\CSIGN$s are only applied
transversely between different encoded blocks, one should choose to 
perform syndrome measurements on the block with fewer erasures, since 
one is less likely to induce a failure that way. In our worst case
approximation, however, we ignore this classical correlation between the
two encoded blocks of data, and we will restrict ourselves to the
partial description of the two qubit output.
Thus, taking the combined output to the $\CSIGN$ to be $\rho$, and
the control and target states to be the result of a partial trace over
$\rho$, that is $\rho_{\text{control}}=tr_{\text{target}}\rho$ and
$\rho_{\text{target}}=tr_{\text{control}}\rho$, 
our error model can be seen as
each type of failure inducing independent superoperators on the qubits
as follows
\begin{subequations}
\begin{eqnarray}
\text{control failure} &:& \left\{
\begin{array}{r@{\to}l}
\rho_{\text{control}}&\er{E}(\rho_{\text{control}})\\
\rho_{\text{target}}&\er{Z}(\rho_{\text{target}}).
\end{array}
\right.\\
\text{target failure}  &:& \left\{
\begin{array}{r@{\to}l}
\rho_{\text{control}}&\er{Z}(\rho_{\text{control}})\\
\rho_{\text{target}}&\er{E}(\rho_{\text{target}}).
\end{array}\right.
\end{eqnarray}
\end{subequations}
Again, we assume that the probability that a photon loss failure has occurred
in either teleportation is independent of the probability that a failure has occurred
in the other teleportation. Moreover, we assume that {\em only} failures due to
photon loss occur in this model, and we assume the probability of 
photon loss in a teleportation is given by $\e_l$ under this model\footnote{The subscript $l$ standing for ``lossy.''}.
\subsection{A mixed model}
In general, depending on the gate construction, the probabilities
of ideal teleportation failure and photon loss determines the error operator for failure of the qubit. 
However, in order to place a bound on the accuracy threshold for
{\em any} gate construction, we consider only the end cases, $\Pr(\text{ideal failure}|\text{failure})=1$ 
and $\Pr(\text{photon loss failure}|\text{failure})=1$, which are the cases described above.
It is important to emphasize that {\em the two error models are very different} -- in one
case we have $\Z$ measurements with known outcomes occuring independently at either qubit of the $\CSIGN$,
while in the other we have, under our worst case approximation, both full and $\Z$ erasure occuring
at either qubits of the $\CSIGN$. The actual error model will be a probabilistic mixture of the
two models considered here.

We can restrict ourselves to considering only the end cases because, 
as will be demonstrated in later sections, the erasures that
follow from photon loss
have a higher cost than the $\Z$ measurements of the ideal model (i.e. more teleported
gates need to be applied),
so the threshold for any mixed error model will fall between the thresholds of these
two end cases. This also provides further motivation for using essentially the same correction procedure
to correct both $\Z$ measurements and $\Z$ erasures, since it simplifies and reduces the
circuitry significantly.

%% file: Candidate-Codes.tex
\chapter{Candidate Codes\label{ch:candidates}}
\markright{Candidate Codes}

In this chapter a comparison is drawn between two small CSS codes, one
being the Steane code, a $[[7,1,3]]$ code often used in threshold calculations
for general errors, and the Grassl code, a $[[4,2,2]]$ code that is the 
smallest single erasure
correcting code. Brief descriptions of universal sets of gates, as well
as details of the erasure correction procedure, are given in order to
justify the preference for the Steane code.

\section{Steane Code\label{sec:steane}}
The Steane code \cite{steane:1996,bank-shor:1996} was one of the first
CSS codes to be discovered. Is it based on the $[7,4,3]$ classical
Hamming code and its dual, the $[7,3,4]$ Simplex code, yielding a
$[[7,1,3]]$ quantum code that allows for very simple fault-tolerant
computation. The 7-bit Hamming code has parity check matrix
\begin{equation}
H_{[7,4,3]}=\left[
\begin{array}{ccccccc}
1 &1 &1 &1 &0 &0 &0\\
1 &1 &0 &0 &1 &1 &0\\
1 &0 &1 &0 &1 &0 &1  
\end{array}\right],
\end{equation}
so following the CSS construction of self-orthogonal classical
codes in Section \ref{subsec:css-as-stab},  
we find that the stabilizer $S$ is generated by 
\begin{equation}\label{eqn:steane-stabs}
\begin{array}{ccccccccccccccl}
\op{M_1}&=&\X&\ox&\X&\ox&\X&\ox&\X&\ox&\I&\ox&\I&\ox&\I\\
\op{M_2}&=&\X&\ox&\X&\ox&\I&\ox&\I&\ox&\X&\ox&\X&\ox&\I\\
\op{M_3}&=&\X&\ox&\I&\ox&\X&\ox&\I&\ox&\X&\ox&\I&\ox&\X\\ 
\op{M_4}&=&\Z&\ox&\Z&\ox&\Z&\ox&\Z&\ox&\I&\ox&\I&\ox&\I\\
\op{M_5}&=&\Z&\ox&\Z&\ox&\I&\ox&\I&\ox&\Z&\ox&\Z&\ox&\I\\
\op{M_6}&=&\Z&\ox&\I&\ox&\Z&\ox&\I&\ox&\Z&\ox&\I&\ox&\Z,  
\end{array}
\end{equation}
Because the structure of the stabilizer generators is directly derived 
from the parity check matrix, it immediately follows that the derived
CSS code has the same minimum distance as the classical Hamming code, as
argued in Section \ref{subsec:css-as-stab}.
\subsection{Fault-tolerant Universal Gates\label{subsec:steane-ft-gate}}
Being a self-orthogonal CSS code, the Steane code has a very simple
fault-tolerant implementation of encoded gates. Shor \cite{shor:1996} was the first 
to demonstrate how universal fault-tolerant computation could be 
performed on the Steane code by explicitly constructing a fault-tolerant
encoded Toffoli using only measurements and encoded Clifford gate operations.
We follow his approach by first giving the encoded Clifford gates demonstrated
by Gottesman \cite{gottesman:1998}, and then use insights by Zhou \etal\ \cite{zlc:2000} to demonstrate
how a generating set for the Clifford group $\C_2$ can be constructed.

First, we need to implement the Pauli operators. Following the
stabilizer formalism, we can choose members of the
normalizer $N(S)$ of the stabilizer that obey the commutation relations for the
Pauli operators, namely:
\begin{subequations}
\begin{eqnarray}
\{\enc{\X},\enc{\Z}\} = 0\\
\enc{\X}\enc{\Z}=\enc{\Y}.
\end{eqnarray}
\end{subequations}
This choice of encoded Pauli operators is equivalent to choosing the
encoded basis states, and it is equally non-unique. One such choice is:
\begin{subequations}
\begin{eqnarray}
\enc{\X} &=& \I\ox\I\ox\I\ox\I\ox\X\ox\X\ox\X\\
\enc{\Z} &=& \I\ox\I\ox\I\ox\I\ox\Z\ox\Z\ox\Z,\label{eqn:enc-z}
\end{eqnarray}
\end{subequations}
but one could just as well call the first operator $\enc{\Z}$ and the second
one $\enc{\X}$, and the computational basis  would
be changed to whatever the eigenbasis of $\enc{\Z}$ is.
These operations do not require interaction between different qubits so they
are automatically fault-tolerant, and this is always the case for encoded Pauli
operations based on the normalizer of the stabilizer code.

The next step is to show the construction of the encoded Clifford gates not in
$\Pauli_k$. We have seen that the Hadamard gate maps between $\X$ and $\Z$ and
leaves the other Pauli operators invariant, and, by construction of the Steane code,
every stabilizer operator that is made up of
only $\X$s has a counterpart that is made up of only $\Z$s, so applying
the the Hadamard transversally preserves the stabilizer, and therefore it
preserves the codespace. Given our choice of $\enc{\X}$ and $\enc{\Z}$,
it is also clear that this operation is the encoded Hadamard itself.
The phase gate $\Pha$
can similarly be applied qubitwise to map
the all $\X$s stabilizer generators into all $\Y$s 
operators with an added phase factor, 
but this is only a valid encoded operation if these 
operators made up of only $\Y$s
are in the stabilizer. Given that
\begin{subequations}
\begin{eqnarray}
\Pha\X\Pha^{\dagger}&=&i\Y\\
\Pha\Z\Pha^{\dagger}&=&\Z,
\end{eqnarray}
\end{subequations}
we would like the weight of the all $\X$s and all $\Z$s 
stabilizer generators be a multiple
of four so that the $\frac{\pi}{2}$ complex phases add up to a $0$ phase. 
Fortunately, the Steane code is a doubly-even CSS code, 
so all stabilizer generators have
weight four. The operation that this qubitwise $\Pha$ performs 
on the encoded operators is
\begin{subequations}
\begin{eqnarray}
\Pha^{\ox 7}\enc{\X}\left(\Pha^{\dagger}\right)^{\ox 7}&=&\I\ox\I\ox\I\ox\I\ox i\Y\ox i\Y\ox i\Y\\
&=&-i\I\ox\I\ox\I\ox\I\ox\Y\ox\Y\ox\Y\\
&=&-i\enc{\Y} = -i\enc{\X\Z}\\
&=&\enc{\Pha}^{\dagger}\enc{\X}\enc{\Pha},
\end{eqnarray}
\end{subequations}
The qubitwise phase gate realizes the encoded inverse phase gate. In order to
get the encoded phase gate itself, one only needs to apply the inverse phase
gate qubitwise.

These constructions of the single qubit Clifford group operations for the Steane code
are much more robust and desirable than the constructions for the same gates for
the Grassl code derivatives, as will be shown later. The main reason for this robustness is the fact that
no two qubit interactions were necessary for any of these constructions, only single
qubit operations. In the context of the KLM proposal for linear optics quantum
computing, this is a powerful advantage, because it allows us to take these operations
to be erasure free in the ideal model, and loss free in our simple lossy model.

\begin{figure}
\centering
\subfigure[Hadamard Gate]{
\includegraphics{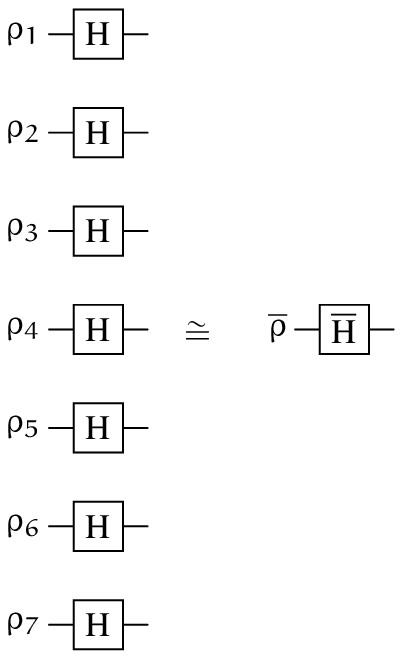}
}
\hspace{1cm}
\subfigure[phase gate]{ 
\includegraphics{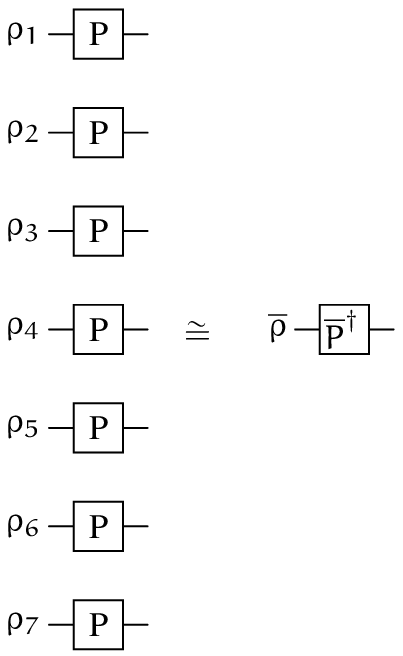}
}
\caption{Transversal one qubit Clifford gates for the Steane code.}
\end{figure}

All we need to complete the generating gates of $\C_2$ is the encoded $\CNOT$. 
For the construction of the encoded $\CNOT$ itself, it is unavoidable that two 
qubit interactions be present. All CSS codes allow for transversal $\CNOT$ 
application as
an encoded operation, and in all these cases the encoded operation that is applied is
itself the encoded $\CNOT$ \cite{gottesman:1998}. This follows again from the fact that the 
stabilizer generators
are made up of either $\X$s or $\Z$s but not both, and following how the $\CNOT$
maps between Pauli operators, it becomes clear that the mapping will preserve
$S\times S$\footnote{We need to consider mappings that 
preserve the set $S \times S$
because the $\CNOT$ is a two qubit mapping, while $S$ describes the 
encoding of only one qubit.}.
A quick calculation demonstrates that 
\begin{subequations}
\begin{eqnarray}
\enc{\X}\ox\I&\to&\enc{\X}\ox\enc{\X} \\
\enc{\Z}\ox\I&\to&\enc{\Z}\ox\I \\
\I\ox\enc{\X}&\to&\I\ox\enc{\X} \\
\I\ox\enc{\Z}&\to&\enc{\Z}\ox\enc{\Z},
\end{eqnarray}
\end{subequations}
as claimed.

\begin{figure}
\centering
\includegraphics{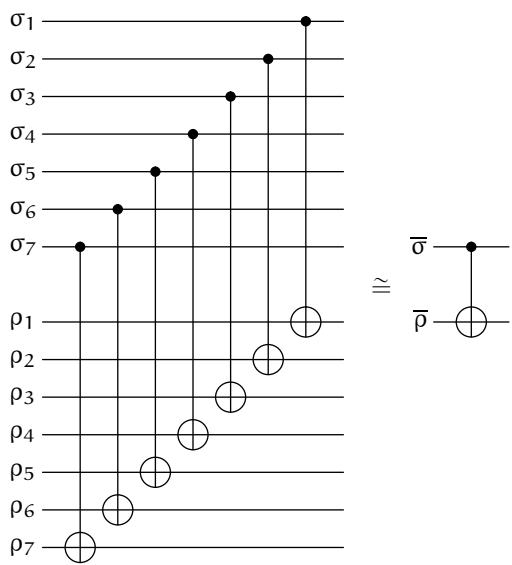}
\caption{Encoded $\CNOT$ for the Steane code.}
\end{figure}

The Clifford group is not sufficient for universal quantum computation. 
However, almost any additional gate that lies outside
$\C_2$ gives us a universal set, and the gates
that are usually considered are the Toffoli gate\footnote{This is the gate
mapping $\ket{a}\ket{b}\ket{c}\to\ket{a}\ket{b}\ket{c\oplus ab}$. It is
also known as the controlled-controlled-not.}
or the $\frac{\pi}{8}$
gate\footnote{This is the gate that leaves the $\ket{\logic{0}}$ unchanged
and maps $\ket{\logic{1}}\to e^{i\frac{\pi}{4}}\ket{\logic{1}}$. It is called
a $\frac{\pi}{8}$ gate for historical reasons.}, which are known to give a universal set when combined
with the Clifford group. The first fault-tolerant construction of
the Toffoli for the Steane code was given by Shor \cite{shor:1996},
and a fault-tolerant construction of the $\frac{\pi}{8}$
gate
was given by Boykin \etal\ \cite{boykin:1999}. Neither of these
constructions was seen to follow the same general framework, but
Zhou \etal\ \cite{zlc:2000} demonstrated how they follow directly from
an extension of the work by Gottesman and Chuang \cite{gottesman-chuang:1999} 
on universal quantum computation using teleportation and single
qubit operations.

\begin{figure}
\centering
\subfigure[$\frac{\pi}{8}$ gate]{
\includegraphics{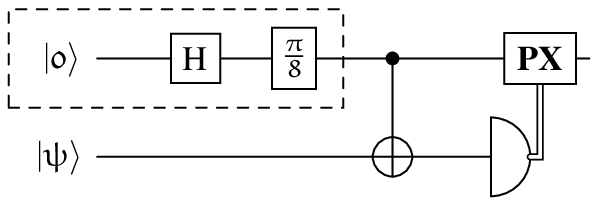}
}
\subfigure[Toffoli gate]{
\includegraphics[scale=0.7]{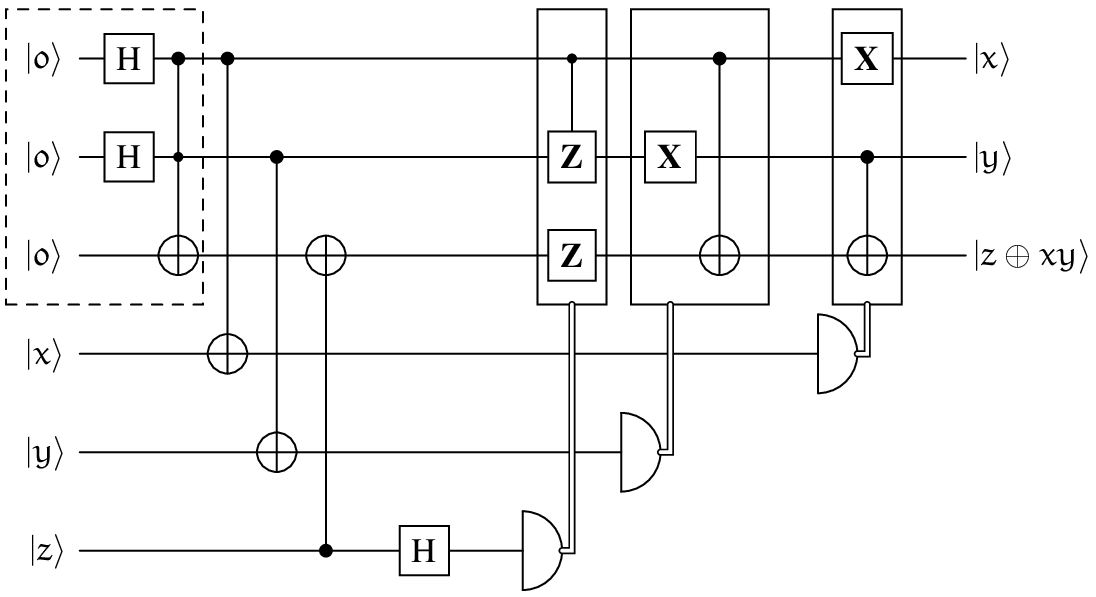}
}
\caption[Fault-tolerant gates needed for universal computation.]{Fault-tolerant gates needed for universal computation, as depicted in \protect\cite{zlc:2000}. 
All measurements are in the $\Z$ eigenbasis.
The dashed boxes represent
states that can be prepared off-line fault-tolerantly, the double lines
represent classical control dependent on the measurement outcome. \label{fig:steane-enc-t}}
\end{figure}

The detailed analysis of how these constructions were obtained
will not be given here, but for illustration the resulting circuits
implementing the $\frac{\pi}{8}$ gate 
and the Toffoli gate are given in Figure
\ref{fig:steane-enc-t}. Zhou \etal\ show how
to prepare the state in the dotted box fault-tolerantly
\footnote{Fault-tolerant state preparation under an erasure model is also easier than
in a general error model, since it is always clear when an error might have occurred}. The construction
of the Toffoli involves not only the three required $\CNOT$s used to perform 
a simplified teleportation protocol,
but also classically controlled application of $\CNOT$s and
$\CSIGN$s depending on the measurement outcomes. This Toffoli
construction still has a threshold very close to the Clifford gate threshold
when the state preparation yield similar failure probabilities as the 
Clifford gate failure probabilities \cite{gottesman-thesis:1997}. 

Given our error model, where $\CNOT$s/$\CSIGN$s are costly
 compared to single qubit gates, it can be intuitively seen that
the encoded $\frac{\pi}{8}$ gate should give a threshold much closer
to the Clifford gate threshold under the same conditions because it
requires fewer applications of such gates.  This will be the operating
assumption for the rest of the analysis, and from now on we focus only
on the threshold for Clifford gates.

\subsection{Erasure Correction\label{subsec:steane-er-corr}}
The Steane code is a $[[7,1,3]]$ quantum code, so immediately
it is clear that it will be able to correct all weight two and one 
erasure patterns. This can be done by measuring the generators of the 
stabilizer of the code
and inferring the erasure operators from the commutation properties
measured -- the collection of all measurements is known as
{\em the syndrome} of the error.

One can check explicitly which erasure patterns are correctable
by applying the modified Knill-Laflamme condition to all
possible erasure operators for a given erasure pattern. Taking the
approach of checking the correctability of $\X$ erasure and $\Z$ erasures
separately simplifies the analysis somewhat -- we can say that
a pattern is correctable if the $\X$ erasure patterns and the
$\Z$ erasure pattern are correctable. Through
this brute force approach, one finds that 
only $28$ of the possible ${7 \choose 3}=35$ weight three patterns, $\frac{4}{5}$ 
of the total, are correctable.

All stabilizer generators have weight four, so any stabilizer measurement
can lead at most to failures in four qubits. If we would like to
measure the syndrome of a single erasure, we simply choose
stabilizers which act non-trivially over the qubit that has been
erased. If there are two erasures, it turns out that one
can always choose a stabilizer that will act non-trivially on
either of the erasures but not on the other one \cite{zalka:personal}.
If we consider single failures on any of the erasure free qubits being
touched by the stabilizer measurement, we find that $\frac{2}{3}$
of all reachable weight three erasure patterns are correctable, but
the other $\frac{1}{3}$ are uncorrectable. This is always the
case, regardless of which of the two erasures we choose to correct,
or which stabilizer we choose to measure (given the constraint that
we do want to correct one erasure when we measure a stabilizer
operator). 

Depending on the type of erasure, a different number of
stabilizer operators need to be measured. In particular, since a $\Z$
erasure just needs to be checked for commutation against an $\X$ stabilizer,
and a $\X$ erasure just against a $\Z$ stabilizer, it follows that a full
erasure needs two stabilizer measurements.

Exactly how the correctability of these erasure patterns can be easily determined,
as well as which patterns are reachable after the erasure correcting
procedure is applied to a given pattern, will be elucidated in Chapter \ref{ch:erasure-threshold}.

\subsection{Z Measurement Correction\label{subsec:z-measurement}}
As mentioned in the previous chapter, we would like to avoid using $\CNOT$s
to measure the $\X$ stabilizer of a CSS code. An alternative method
has been developed that allows for measurement of the syndrome and correction
of the error to be done in a single step, as long as the outcome of
the $\Z$ measurement is available \cite{manny:2001,klm-thr:2000,klm:2001}.

\begin{figure}
\centering
\includegraphics{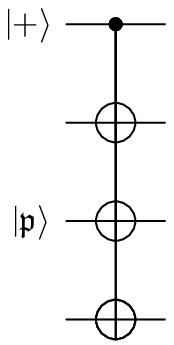}
\caption[Circuit taking the state $\ket{\logic{p}}$ into an eigenvalue $+1$
eigenvector of $\X^{\ox 4}$.]{Circuit taking the state $\ket{\logic{p}}$ into an eigenvalue $+1$
eigenvector of $\X\ox\X\ox\X\ox\X$.\label{fig:extend-and-correct}}
\end{figure}

In order to correct the $\Z$ measurement, we need to, indirectly,
 measure an $\X$ stabilizer that acts non-trivially over a single
$\Z$ measurement. Since the Steane code only has weight four $\X$ stabilizers,
we can consider only the four qubits over which the stabilizer acts 
non-trivially, so the stabilizer can be rewritten
\begin{equation}
\X\ox\X\ox\X\ox\X,
\end{equation}
with $+1$ eigenvalue eigenvectors
\begin{eqnarray}
\frac{1}{\sqrt{2}}(\ket{\logic{0000}}+\ket{\logic{1111}})&=&\ket{\gamma_1}\\
\frac{1}{\sqrt{2}}(\ket{\logic{0101}}+\ket{\logic{1010}})&=&\ket{\gamma_2}\\
\frac{1}{\sqrt{2}}(\ket{\logic{0011}}+\ket{\logic{1100}})&=&\ket{\gamma_3}\\
\frac{1}{\sqrt{2}}(\ket{\logic{0110}}+\ket{\logic{1001}})&=&\ket{\gamma_4}.
\end{eqnarray}
Without loss of generality, take the first qubit to be the qubit that has
undergone a $\Z$ measurement, and we apply an $\X$ on all remaining qubits
if the outcome is $\ket{\logic{1}}$, and do nothing otherwise. 
If initially we have some state
\begin{equation}
\sum_{i=1}^4\alpha_i\ket{\gamma_i},
\end{equation}
then after the measurement (and possible
$\X$ applications) we will have the state
\begin{equation}
\ket{\logic{p}}=\sum_{i=1}^4\alpha_i\ket{\tilde{\gamma}_i},
\end{equation}
where, if we consider only the non-measured qubits, we have
\begin{eqnarray}
\ket{\tilde{\gamma}_1}&=&\ket{\logic{000}}\\
\ket{\tilde{\gamma}_2}&=&\ket{\logic{101}}\\
\ket{\tilde{\gamma}_3}&=&\ket{\logic{011}}\\
\ket{\tilde{\gamma}_4}&=&\ket{\logic{110}}.
\end{eqnarray}
In order to obtain the original state back, we need a fresh qubit
in the state $\ket{+}=\frac{1}{\sqrt{2}}\left(\ket{\logic{0}}+\ket{\logic{1}}\right)$,
so we apply the circuit in Figure \ref{fig:extend-and-correct}. Since this
circuit is made up of only $\C_2$ gates, we can consider the teleportation
of the state $\ket{\logic{p}}$ before applying Figure \ref{fig:extend-and-correct},
and then simply commute that part of the circuit into the state preparation part
of the teleportation protocol -- the only qubit introduced is in the
fixed state $\ket{+}$, so it can be taken into the state preparation part as well.

\begin{figure}
\centering
\includegraphics{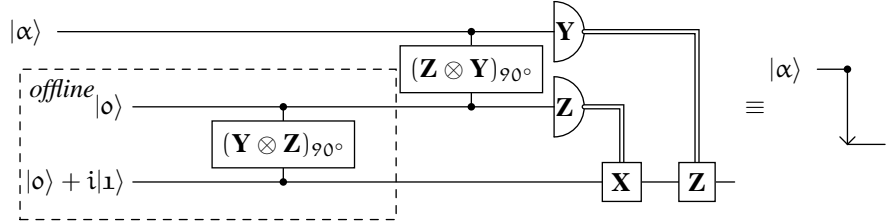}
\caption[Modified teleportation protocol.]{Modified teleportation protocol introducing only $\Z$ measurements
in case of the failure of the $(\Z\ox\Y)_{90^\degg}$ gate \label{fig:z-measurement-teleport}}
\end{figure}

It is important that we use the slightly modified teleportation protocol
shown in Figure \ref{fig:z-measurement-teleport}. In this qubit teleportation protocol,
the only source of failure is the two qubit gate $(\Z\Y)_{90^\circ}$, which is
performed through a $\CSIGN$ conjugated by one qubit $\C_2$ gates (the other two qubit
gates are performed similarly). This teleportation protocol guarantees that if there 
is a failure in the $(\Z\Y)_{90^\degg}$ gate, it will translate to a $\Z$ measurement
of the qubit being teleported. Because of the structure of the circuit in
Figure \ref{fig:extend-and-correct}, only the $\Z$ corrections of the teleportation will
propagate through the $\CNOT$s, and therefore so will the $\Z$ measurements.

\begin{figure}
\centering
\includegraphics{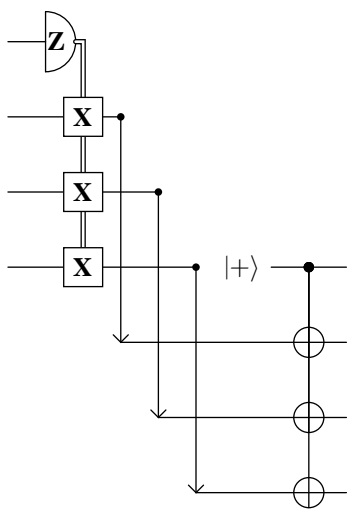}
\caption[Full circuit for correction of a single $\Z$ measurement.]{Full circuit for correction of a single $\Z$ measurement. The $\CNOT$s can be commuted
backwards in time and combined with the offline state preparation for the teleportations.}
\end{figure}

If any of the teleportations fail, the corresponding qubit undergoes a
$\Z$ measurement, but also the correction target, the qubit we were trying to
recover, cannot be corrected. Taking each teleportation to fail with
a probability $\e_i$, the correction will succeed with probability
$(1-\e_i)^3$, and if we take $\er{Z}$ to signify the $\Z$ measurement,
we have the following transition probabilities on the four qubits
in question
\begin{eqnarray}
\Pr(\I\ox\I\ox\I\ox\I | \er{Z}\ox\I\ox\I\ox\I)&=&(1-\e_i)^3\\
\Pr(\er{Z}\ox\er{Z}\ox\I\ox\I | \er{Z}\ox\I\ox\I\ox\I)&=&\e_i(1-\e_i)^2\\
\Pr(\er{Z}\ox\er{Z}\ox\er{Z}\ox\I | \er{Z}\ox\I\ox\I\ox\I)&=&\e_i^2(1-\e_i)\\
\Pr(\er{Z}\ox\er{Z}\ox\er{Z}\ox\er{Z} | \er{Z}\ox\I\ox\I\ox\I)&=&\e_i^3
\end{eqnarray}
with the other patterns with equal weights being obtained
by permutation, preserving the probabilities of transition.

In this analysis, we have neglected the remaining three qubits of the
Steane code, and we have assumed that the state described by the four
qubits in question was a pure state, which is not true. In reality,
the state of those 4 qubits will be a mixture, because the seven qubits
of the code block are entangled, but similar results in that case follow
by linearity from the results given here.

With the method described here, we can correct $\Z$ measurements
without directly measuring $\X$ stabilizers. We can employ the exact
same method to correct $\Z$ erasures by first measuring the
erased qubit in the $\Z$ basis, and then simply replacing it
with a qubit corresponding to the measurement outcome. 
Once the measurement outcome is known, this
same procedure can be used.

In the case of a qubit loss in any of the three teleportations, we take
the worst case approach once more. Thus, the qubit being teleported
is completely erased, and the $\Z$ erased qubit that is being corrected
is $\Z$ erased once more. Each of the three teleportations can lose a photon
independently, so the transition probabilities can easily be computed
for the lossy model as well, yielding
\begin{eqnarray}
\Pr(\I\ox\I\ox\I\ox\I | \er{Z}\ox\I\ox\I\ox\I)&=&(1-\dt)(1-\e_l)^3\\
\Pr(\er{E}\ox\I\ox\I\ox\I | \er{Z}\ox\I\ox\I\ox\I)&=&\dt\\
\Pr(\Z\ox\er{E}\ox\I\ox\I | \er{Z}\ox\I\ox\I\ox\I)&=&(1-\dt)\e_l(1-\e_l)^2\\
\Pr(\Z\ox\er{E}\ox\er{E}\ox\I | \er{Z}\ox\I\ox\I\ox\I)&=&(1-\dt)\e_l^2(1-\e_l)\\
\Pr(\Z\ox\er{E}\ox\er{E}\ox\er{E} | \er{Z}\ox\I\ox\I\ox\I)&=&(1-\dt)\e_l^3,
\end{eqnarray}
with the factors involving $\dt$, the probability of a detector failing,
arising due to the fact that we need
to perform a $\Z$ measurement explicitly.
\subsection{The Encoded Error Model\label{subsec:encoded-error-model}}
In the case of a failure in correction an error in a code block, it is
important to understand what kind of error is induced in the encoded
qubit. This encoded error model is crucial for predicting
the performance when the information is encoded multiple
times, an approach that will be used in the Chapter \ref{ch:erasure-threshold}.

For the error models considered in Chapter \ref{ch:eloqc}, this is a relatively
simple problem that can be solved with the help of the stabilizer and normalizer 
formalisms.

\subsubsection{Error Model for Ideal Hardware}
In this error model, all erasures are $\Z$ measurements, and from the
structure of the Steane code, it is known which patterns of erasures are
correctable and which ones are not. Consider the case of the $\Z$ measurement
of the last 3 qubits of a code block, which is not correctable. 
Since the encoded $\Z$ operation
consists of $\Z$s applied to the last 3 qubits, as described by \eqref{eqn:enc-z},
it is clear that this uncorrectable error is equivalent to an encoded
$\Z$ measurement on the encoded qubit. Consider, on the other hand,
the case of $\Z$ measurements on all qubits of the code block. Since,
given the stabilizer operator $\op{M}_4$ from \eqref{eqn:steane-stabs},
$\enc{\Z}\op{M}_4=\Z\ox\Z\ox\Z\ox\Z\ox\Z\ox\Z\ox\Z$, it is clear that
this is also equivalent to a $\enc{\Z}$ measurement. It turns out that
because of this, all uncorrectable weight three measurement patterns (as well
as the weight seven pattern) are  equivalent to a $\enc{\Z}$ measurement.
All uncorrectable measurement patterns of intermediate weight can be thought of
as a uncorrectable weight three measurement pattern followed by some
additional measurements, so that all of them are equivalent to a $\enc{\Z}$
measurement.

In practice, it is also important to consider the cases of correctable
erasure patterns, since a code block may end up in such a state after
some finite number of error correcting rounds. This is extremely simple
for the case of ideal hardware, since we can simply measure any number
of qubits perfectly, allowing us to force an uncorrectable measurement
pattern, which as shown above is always a $\enc{\Z}$ measurement.

Thus, the encoded error model is identical to the base error model
up to different probability distributions.

\subsubsection{Error Model for Lossy Detectors}
Given the correspondence between $\Z$ erasures and $\Z$ measurements
of unknown outcome\footnote{See \eqref{eqn:z-meas-outcome} in Chapter \ref{ch:eloqc}.},
failures consisting only of $\Z$ erasures have a behavior that parallels
failures consisting only of $\Z$ measurements. 

The stabilizer generators for the Steane code that are tensor products of $\X$s only
have exactly the same structure as the stabilizer generators that are
tensor products of $\Z$s only, and a full erasure is the composition
of a $\Z$ erasure and an $\X$ erasure\footnote{The order of which partial
erasure occurs first is irrelevant.}, so uncorrectable 
erasure patterns consisting only of full erasures are equivalent to
encoded full erasures.

Uncorrectable erasure patterns that include both $\Z$ and full erasures
need to be considered as composition of $\Z$ and $\X$ erasures patterns.
With the error correction scheme described in the previous sections of this
chapter, we can only have $\X$ erasures on qubits that have $\Z$ erasures 
as well -- that is to say that we only have $\Z$ or full erasures, but
no $\X$ erasures. If the $\X$ erasure pattern is uncorrectable, then
the encoded failure will be an encoded full erasure, since we are guaranteed
to also have an uncorrectable $\Z$ erasure pattern. On the other hand, if
the $\X$ erasure pattern is correctable, we can simply attempt to
correct $\X$ erasures until either there are no more full erasures (in which case we would
only have the uncorrectable $\Z$ erasure pattern, which corresponding
to an encoded $\Z$ erasure), or until the $\X$ erasure pattern is uncorrectable
(in which case we would have the encoded full erasure).

Correctable erasure patterns are a little more complex since they
require the measurement of all qubits in order to be interpreted as encoded
errors, and because of the lossy nature of the detectors, the outcome
could be either an encoded $\Z$ measurement, or an encoded full erasure. The
probabilities of each of these outcomes depends on the probability of failure of
the qubit measurements, so this extra measurement needs to be regarded as a separate step
after the desired rounds of error correction. In the best case, where no
measurements fail, we obtain an encoded $\Z$ measurement, while in the worst case,
we obtain an encoded full erasure.
As shown before, measuring all qubits in the $\Z$ eigenbasis is equivalent to
an encoded $\Z$ measurement. Moreover, because of the structure of
the Steane code, even though some of these measurements may fail, the
encoded measurement outcome may still be inferred by applying classical
error correction techniques to the outcomes. If recovery is not possible
because of the high number of errors, then the failure is an encoded
full erasure. We can take these
correctable erasure patterns to be encoded $\Z$ erasures, since, like
$\Z$ erasures, they require $\Z$ measurements to be performed as the first step
of the error correction, and failure of the measurement yields a full erasure.

Once again, the encoded error model is identical to the base error model
up to a different probability distribution.

\subsubsection{A Mixed Error Model}
In a mixed error model, where we have both the limitations of linear optics
and the post-selection based construction of the $\CSIGN$ as well as the
imperfect photon detectors, the error model changes only quantitatively.
At higher levels of encoding we still expect to find $\Z$ measurements
as well as $\Z$ and full erasures, but qualitative analysis shows that
there is a bias towards the more benign error model.

A failure pattern than includes both $\Z$ erasures and $\Z$ measurements
(but no full erasures)
can be taken as an encoded $\Z$ erasure at no additional cost, or
may be taken as an encoded $\Z$ measurement at the cost of measuring
the $\Z$ erasures, possibly introducing full erasures which may lead
to encoded full erasures. If the measurement can be inferred without
measuring the $\Z$ erasures by using classical error correction, 
the probability of encoded full erasures is reduced. Failure patterns
that include full erasures are treated similarly.
  
The procedure for turning correctable failures into
failures at a higher level of encoding could either yield $\Z$ measurements, full erasures 
or $\Z$ erasures. In particular, there is no need to explicitly measure
qubits that have been affected by an unintentional measurement, which also 
reduces the probability of encoded full erasures.

Both these opportunities for gain over the purely lossy error model indicate
that the mixed error model should always allow for improvement over the
purely lossy error model, and that it will be naturally biased towards the
ideal detector error model.

\subsubsection{Worst Case Analysis}

For the calculations made in Chapters \ref{ch:erasure-threshold}, we take
all encoded errors in the error model for lossy detectors 
to be full erasures. This worst case approach simplifies
the calculations significantly, and although it gives a looser lower 
bound to the error threshold, it will be shown to be enough to match
predictions in \cite{klm-thr:2000, klm:2001}.

Formally one usually takes the threshold to be the break even point
between the encoded and the base error rates. Here we will take the
threshold to be the point where the encoded error rate is half of the
base error rate. This is because the probability of full erasure of 
a qubit given a failure has occured is the same as the probability of
a $\Z$ erasure given a failure has occured -- that is, both occur with
conditional probability $\frac{1}{2}$. Thus, since we take all encoded 
failures to be full erasures, we apply the break even condition to
full erasures only, which translates to the requirement that the overall
encoded failure rate be half of the base failure rate. When the threshold
is explicitly calculated in Section \ref{subsec:calc}, this relationship
will be formalized more clearly.

\section{Grassl Code}
Grassl, Beth and Pellizzari \cite{grass-etal:1997} have shown that
the shortest possible erasure code is a $[[4,2,2]]$ code, which is referred
to here as {\em the Grassl code}. This is a CSS code derived from
the classical code consisting of all even weight strings of length
four. It corrects a single erasure, the general class of such distance two
codes having been discussed in detail in
\cite{gottesman:1998}. The stabilizer group $S$ of this code is generated by the
operators
\begin{subequations}
\begin{eqnarray}
\begin{array}{ccccccccc}
\op{M}_1&=&\X&\ox&\X&\ox&\X&\ox&\X\\
\op{M}_2&=&\Z&\ox&\Z&\ox&\Z&\ox&\Z,
\end{array}
\end{eqnarray}
\end{subequations}
and a generating set for the encoded Pauli operators is
\begin{subequations}
\begin{eqnarray}
\begin{array}{cccccccc}
\enc{\X}_1 = &\X&\ox&\X&\ox&\I&\ox&\I \\
\enc{\X}_2 = &\X&\ox&\I&\ox&\I&\ox&\X \\
\enc{\Z}_1 = &\Z&\ox&\I&\ox&\I&\ox&\Z \\
\enc{\Z}_2 = &\Z&\ox&\Z&\ox&\I&\ox&\I,
\end{array}
\end{eqnarray}
\end{subequations}
The fact that all operators in the generating set of the encoded Pauli
operators are of weight two proves that this is a distance two code --
that is, there are weight two error operators that will map valid codewords
into different valid codewords, and will therefore be undetectable and
uncorrectable.

\subsection{Fault-tolerant Universal Gates}
One difference between the Grassl code and the Steane code that is immediately
obvious is the fact that the Grassl code encodes two qubits. This raises
the question of how to concatenate the code, and how the different choices affect
the performance. Even if this issue is resolved, because we have two encoded
qubits, we must consider different types of encoded operations dealing with
the different encoded qubits. While there are very simple constructions of
some of these gates -- such as an encoded $\CNOT$ between the two encoded
qubits in a block or the $\SWAP$ between these same qubits --
the construction of other types of $\CNOT$s have a very high
cost because of the use of multiple rounds of teleportation and multiple
ancilla preparations -- some gates that suffer this problem are the
encoded $\CNOT$ between the individual encoded qubits of two different blocks.

On the other hand, one may choose to bypass this problem by modifying
the Grassl code to encode a single qubit. That can be done by choosing
one of the encoded Pauli operators over a single encoded qubit 
and making it a stabilizer. Any such choice is
valid since we know that all operators in $N(S)$ commute with the elements
of $S$ (see Section \ref{subsec:norm-and-heisenberg}), and that
results in a stabilizer code with three stabilizer generators, as desired
($n-k=4-1$ since we want a $[[4,1]]$ code). It is easy to check
that the minimum distance of the code remains the same, so we are able
to correct a single erasure with this code. The main problem is that
the resulting code is not a self-orthogonal CSS code, but a more
general form of CSS code. While the
encoded $\CNOT$ can easily be implemented by qubitwise $\CNOT$s,
neither the encoded Hadamard nor the encoded Phase gates can
be implemented as easily. Instead, it is necessary to use 
ancillary states, multiple $\CNOT$s and single qubit operations
to generate these two remaining encoded Clifford gates \cite{gottesman:1998}.
This is highly undesirable for the error models under
consideration because this translates to encoded single
qubit operations that can introduce erasures, while
in the Steane code those operations can never introduce erasures.

A similar problem is found even if the Grassl code is maintained in
its original form. The encoded single qubit operations that are applied
only to one encoded qubit, while the identity is applied to the other,
require complex protocols involving teleportations and measurements.
Thus, there is strong evidence that the Grassl code is suboptimal
for fault-tolerant quantum computation in the error model given
for linear optics quantum computation.

%% file: Threshold.tex
\chapter{The Erasure Threshold\label{ch:erasure-threshold}}
\markright{The Erasure Threshold}

A novel method for determining the recursion
relation for the encoded failure rate is given. This method
describes the erasure correction procedure as a random walk over 
equivalence classes of erasure patterns, allowing for a compact
description that can give the recursion relation for any
number of erasure correction attempts. Using this technique,
the recursion relations for the Steane code under the ideal
and lossy error models are given, and values for the threshold
are estimated numerically.
\section{Calculation of the Threshold\label{subsec:calc}}
In order to calculate the error threshold, we need to calculate the
recursion relation for the probability of error $\e^{(L)}$ in
encoding level $L$ in terms of the probability of error
$\e^{(L-1)}$ in encoding level $L-1$. For simplicity, 
we consider only the relationship
between the first level of encoding and the base error rate, that is, 
\begin{equation}
\label{eqn:recursion-gen}
\e^{(1)}=\sum_{n=m}^{\infty}c_n\e^n,
\end{equation}
 where $\e$ is the error rate at the physical qubits, $c_n$ are constant
integers dependent only on $n$, and $m$ is the minimum number of failures causing an unrecoverable
error, as
we have seen in Chapter \ref{ch:qecc-ft}.

In the general case one needs to keep track of the possible propagation
of undetected errors, but the erasure error model allows us to
ignore this complication because the presence and location of
erasures is known by definition. Thus, we only apply the erasure correction
procedure when we know an erasure has occurred, and any additional
failure that occurs during erasure correction is immediately flagged, so we can
keep applying the erasure correction procedure until the data is
erasure free or we find the resulting threshold for the number of
trials is acceptable.

Since the Steane code is a $[[7,1,3]]$ code, we know it can correct
any one or two erasure patterns on any given code block, so it is impossible
that only one or two erasures at one level will cause an erasure at a higher
level of encoding. Some weight three erasure patterns are not correctable,
so it is possible for three erasures at one level to cause a failure at a
higher level. Therefore, the leading term of the series expansion of
\eqref{eqn:recursion-gen} will be cubic power in $\e$, i.e.
\begin{equation}
\e^{(1)}=\sum_{n=3}^{\infty}c_n\e^n=c_3\e^3+c_4\e^4+\cdots.
\end{equation}
We cannot restrict the analysis to erasures that occur
during computational steps only, but we also need to consider erasures that are introduced 
during the erasure correction steps, since both are performed in the same
physical system.

In essence, we need to consider erasures that occur during
\begin{itemize}
\item computation
\item interaction with ancillae used for syndrome measurement 
\item measurement of ancillae used for syndrome measurement
\end{itemize}
Once the recursion relation has been computed, the threshold is found
by solving an equation describing the break even point, that is,
the probability $\e$ at which the encoded failure rate is
equal to the base failure rate
\begin{equation}\label{eqn:regular-break-even}
\e^{(1)}=\e,
\end{equation}
which under our simplifying assumption, gives, for $\e^{(1)}<\e$
the required condition $lim_{L\to\infty}\e^{(L)}=0$. Note that for
the lossy detector error model, we modify the break even condition
to be the worst case break even condition, since we take all failures
to lead to encoded full erasures. Moreover,
the probability of a single qubit full erasure is half of the total
probability of failure, yielding the worst case
break even condition
\begin{equation}\label{eqn:worst-case-break-even}
\e_l^{(1)}=\frac{1}{2}\e_l.
\end{equation}
Because the recursion relation describing $\e^{(1)}$ is of very large
(if not infinite) order in $\e$, these equations cannot be solved exactly,
but it can be approximated numerically. It is not clear 
how many leading terms of the recursion
relation are sufficient to accurately determine the error threshold, 
so we would like to be able to easily compute as many of the terms as possible.

One approach is to consider all possible correctable erasure patterns.
After an application of the encoded $\CSIGN$, the only source of failures
in our error model, the erasure pattern determined by the failures
of the individual gate is known. By attempting to correct the erasure,
one applies $\CSIGN$s between the data and the ancilla in order to 
measure stabilizer operators (see Section \ref{subsec:ft-stab-meas}). 
Note that we can restrict ourselves
to $\CSIGN$s because a $\CNOT$ is simply given by
\begin{equation}
\CNOT_{1,2}=(\I\ox\Had)\CSIGN_{1,2}(\I\ox\Had),
\end{equation}
and if we only apply the second Hadamard gate after all erasures
have been corrected, we have the advantage of having an error model
that still consists only of $\er{E}$ (full erasures) and $\er{Z}$ 
(\Z\ erasures). 

Any failure during the erasure correction will change the erasure pattern
in the data into some other erasure pattern. Since we know the probability
of the different failures occurring, and we know which operators will have to
be measured in order to correct a given erasure pattern, we may describe
the process by transition probabilities between different erasure patterns.
In the case that the erasure rate at each time step is independent of the
previous time steps, what ends up being described is a {\em Markov chain},
or a {\em random walk in a graph}. Therefore, given the set of possible
correctable erasure patterns ${\er{W}_i}$, we want to calculate all
transition probabilities $\Pr(\er{W}_i|\er{W}_j)$ for some error model.
If we have a description of the initial distribution of erasure patterns,
we can compute what is the probability of each erasure pattern after
an erasure correcting step by following a simple application
of Bayes' theorem
\begin{equation}
\Pr(\er{W}_i,t_{i+1})=\sum_j \Pr(\er{W}_i|\er{W}_j)\Pr(\er{W}_j, t_i),
\end{equation} 
where we take $t_{i+1}$ to be the time instant immediately following
$t_i$. Once the transition probabilities have been computed, by repeated
application of this formula one obtains the probability distribution
of all erasure patterns at time $t_N$, which corresponds to the
distribution after $N$ rounds of the erasure correction procedure.
At that stage, any erasure pattern that has weight greater than 0
is considered a failure, and if we compute
\begin{equation}
\e^{(1)}=\sum_{\text{wt}(\er{W}_i)>0}\Pr(\er{W}_i,t_N),
\end{equation}
we have the recursion relation for $N$ rounds of erasure correction\footnote{
Note that as $N\to\infty$, the probability distribution converges to a stable
distribution.}.

We may rewrite the above transition probabilities into a
transition matrix ${\mathscr P}$ with elements
\begin{equation}
[{\mathscr P}]_{ij}=\Pr(\er{W}_i|\er{W}_j).
\end{equation}
Then the probability distribution of the erasure patterns may be described by
a column vector ${\mathscr I}(t_0)$ given by
\begin{equation}
[{\mathscr I}(t_0)]_{j}=\Pr(\er{W}_i,t_0).
\end{equation}
This allows for a recursive definition of the probability distribution at time $t_{N}$
\begin{equation}\label{eqn:prob-dist-rule}
{\mathscr I}(t_N)={\mathscr P}{\mathscr I}(t_{N-1})={\mathscr P}^N{\mathscr I}(t_0).
\end{equation}
Because of the way in which ${\mathscr P}$ was defined, it satisfies the conditions
of a {\em stochastic matrix}, which guarantees that ${\mathscr I}(t_N)$ as
defined in \eqref{eqn:prob-dist-rule} is a probability distribution
as long as ${\mathscr I}(t_0)$ is one as well.

It is also worth noting that certain erasure patterns in the above
description have special properties. Consider, for example, the
erasure pattern of weight zero. There is no need to apply the
erasure correcting procedure to this erasure pattern, so the
probability of transition from this pattern to any other of
non-zero weight is zero.  This is what is called an {\em absorbing 
state}. Other examples of absorbing states are uncorrectable
erasure patterns. Since we know that there is no way we can go from
an uncorrectable erasure pattern to a correctable one, there is no
point in applying the erasure correcting procedure either.

This approach, although simple to describe, is by no means efficient.
The number of possible erasure patterns grows exponentially in the
number of qubits in the code.
\subsection{Equivalence of Error Superoperators}
There is great redundancy in the description of the transition
probabilities at each erasure correction step. In the context of
the Steane code, if we observe all transitions from a weight
one erasure pattern, we notice they are very similar. Namely,
they will have the same number of transitions to patterns of
any given weight, and each of these transitions will have the same probability.

\begin{figure}
\centering
\includegraphics[scale=0.5]{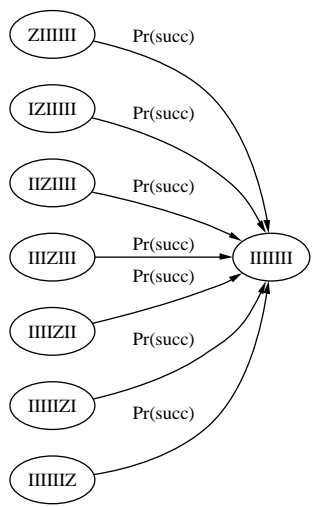}
\caption[Portion of a naive Markov chain describing error correction.]{Portion of a naive Markov chain describing the erasure correction
of some weight one erasure patterns. Here $\Pr(succ)$ indicates the
probability of success in correcting one phase erasure.\label{fig:naive-markov}}
\end{figure}

The regularity of the graph depicted in Figure \ref{fig:naive-markov} 
is no coincidence -- it is, in fact, a
direct consequence of the symmetries of the erasure correction code
being employed. Intuitively, many of these erasure superoperators
are equivalent, and can be `bundled up' into a single representative
superoperators, eliminating the redundant edges and vertices in the
graph in favor of a minimal description that differentiates between
the operational behavior of the superoperators -- that is, their
correctability, and their transitions to other superoperators. 
Formally, we need to describe an equivalence relation between 
erasure superoperators. In order to do that, we introduce the concept of
a code automorphism.

\begin{definition} The automorphism group $Aut(C)$ of a stabilizer code $C$
is defined as the subgroup of the group generated by qubit permutations 
and qubitwise
applications of $\Had$ and $\Pha$ leaving the
stabilizer of $C$ invariant \cite{calderbank-etal:1998}.
\end{definition}

The code automorphism can be used to define the equivalence relation we need.

\begin{definition} For a given quantum code $C$
with automorphism group $Aut(C)$, and given the Pauli group over $n$ 
qubits $\Pauli_n$, we define the relation $\R_C$ over $\Pauli_n$ such that
\begin{equation}
\op{E}\R_C\op{F} \leftrightarrow \exists g\in Aut(C)\ : \op{E}=g\op{F}g^{-1}.
\end{equation}
\end{definition}
Given this definition, some elementary results about $\R_C$ can be given.
\begin{proposition} $\R_C$ is an equivalence relation over $\Pauli_n$.
\end{proposition}
\begin{proof}
Note that $Aut(C)$ is a group, so, by definition, is it closed under 
product, it contains the identity, and every element has an inverse. In the following, take
$\op{E}_i\in \Pauli_n$ for any index $i$. We show that $\R_C$ has each of
the properties of equivalence relations:

{\em Reflexivity:} It is always the case that $\op{E}_i\R_C\op{E}_i$.

This follows trivially from the fact that, because $Aut(C)$ is a group, $\I\in Aut(C)$,
so $g=g^{-1}=\I$ and $\op{E}_i=g\op{E}_ig^{-1}$.

{\em Symmetry:} If $\op{E}_i\R_C\op{E}_j$, then $\op{E}_j\R_C\op{E}_i$.

It is given that $\op{E}_i=g\op{E}_jg^{-1}$, so conjugating both sides by $g^{-1}$ yields
$g^{-1}\op{E}_ig=\op{E}_j$. Since $g^{-1}\in Aut(C)$, it follows that $\op{E}_j\R_C\op{E}_i$

{\em Transitivity:} If $\op{E}_i\R_C\op{E}_j$ and $\op{E}_j\R_C\op{E}_k$, 
then $\op{E}_i\R_C\op{E}_k$.

It is given that $\op{E}_i=g\op{E}_jg^{-1}$ and $\op{E}_i=h\op{E}_kh^{-1}$. By substitution,
$\op{E}_i=gh\op{E}_kh^{-1}g^{-1}$, and since $g,h\in Aut(C)$, it follows that $gh,h^{-1}g^{-1}=(gh)^{-1}\in Aut(C)$
and that $\op{E}_i\R_C\op{E}_k$.

This concludes the proof.
\end{proof}

\begin{theorem}
If error operators $\{\op{E}_i\}$ satisfy the modified Knill-Laflamme condition for
some code $C$, then so do $\{g\op{E}_ig^{\dagger}\}$, where $g\in Aut(C)$.
\end{theorem}
\begin{proof}
The first modified Knill-Laflamme condition states that
\begin{equation}
\bra{c_l}\op{E}_i\ket{c_l}=\bra{c_m}\op{E}_i\ket{c_m}
\end{equation}
for any basis codewords $\ket{c_l},\ket{c_m}\in C$. Since $g \in Aut(C)$, so is
$g^{-1}=g^{\dagger}$, and one can insert the identity $gg^{\dagger}=g^{\dagger}g=\I$ 
in the product $\bra{c_l}\op{E}_i\ket{c_l}$
\begin{subequations}
\begin{eqnarray}
\bra{c_l}\op{E}_i\ket{c_l}&=&\bra{c_l}g^{\dagger}g\op{E}_ig^{\dagger}g\ket{c_l}\\
&=&\bra{\tilde{c}_l}g\op{E}_ig^{\dagger}\ket{\tilde{c}_l}
\end{eqnarray}
\end{subequations}
where the automorphism maps the basis $\{\ket{c_i}\}$ into $\{\ket{\tilde{c}_i}\}$,
and since all code automorphisms are unitary, both bases are orthonormal. Moreover,
both bases describe the same space -- the code space -- so operators
satisfying the Knill-Laflamme conditions in either basis are correctable.
More explicitly, the first condition can be re-written
\begin{equation}
\bra{\tilde{c}_l}g\op{E}_ig^{\dagger}\ket{\tilde{c}_l}=\bra{\tilde{c}_m}g\op{E}_ig^{\dagger}\ket{\tilde{c}_m},
\end{equation}
so clearly it is satisfied for any $\{g\op{E}_ig^{\dagger}\}$ regardless
of the automorphism $g\in Aut(C)$.

The second modified Knill-Laflamme condition states that
\begin{equation}
\bra{c_l}\op{E}_i\ket{c_m}=0\text{ for }\braket{c_l}{c_m}=0.
\end{equation}
Again, it follows quite simply that
\begin{subequations}
\begin{eqnarray}
\bra{c_l}\op{E}_i\ket{c_m}&=&\bra{c_l}g^{\dagger}g\op{E}_ig^{\dagger}g\ket{c_m}\\
&=&\bra{\tilde{c}_l}g\op{E}_ig^{\dagger}\ket{\tilde{c}_m}\\
&=&0,
\end{eqnarray}
\end{subequations}
and that
\begin{subequations}
\begin{eqnarray}
\braket{c_l}{c_m}&=&\bra{c_l}g^{\dagger}g\ket{c_m}\\
&=&\braket{\tilde{c}_l}{\tilde{c}_m}\\
&=&0,
\end{eqnarray}
\end{subequations}
concluding the proof.
\end{proof}
Note that, because $Aut(C)$ is made up
of permutations and tensor products of single qubit unitary operations
in $N(S)$, where $S$ is the stabilizer of the code $C$,
the weight of $g\op{E}_ig^{\dagger}$ is the same as $\op{E}$ for
any $\op{E}\in\Pauli_n$ and $g\in Aut(C)$ for {\em any} code $C$.

It has been shown in Chapter \ref{ch:qecc-ft} that an erasure pattern is a uniform, convex sum over
Pauli erasure operators. In that case, we can extend the definition of
$\R_C$ to be a relation over erasure patterns as well. 
We say $\er{W}\R_C\er{K}$ if and only if there is a $g\in Aut(C)$ mapping between the set of all erasure
operators $\left\{\op{E}_i\right\}_{i}$ in the convex sum $\er{W}$ to the set of all erasure operators
$\left\{\op{F}_j\right\}_{j}$ in the convex sum $\er{K}$. 

In some cases, however,
we want to change which operations $g$ are allowed in this definition because the error correction
procedure may not be invariant under all code automorphisms. 
The problem becomes clear if we consider an error model that 
allows for $\Y$ erasures or $\X$ erasures, which under this definition are
equivalent to $\Z$ erasures.
The $\Y$ erasure requires
two error correction steps in the best case and thus has different behaviour from a
 $\Z$. Similarly, we have used different circuitry to correct $\X$ erasures and thus, if $\X$
erasures were not always accompanied by $\Z$ erasures, they should not be 
considered equivalent to $\Z$ erasures. For the error model considered here,
this distinction is not necessary, and the definition in term of the automorphism group
is suitable and simple to analyze. However, in general, we want the equivalence relation
to relate erasure patterns that have equivalent error correction behaviour, and a different
error model may require a different definition for the equivalence relation between
the erasure patterns. A simple restriction, for example, would be to consider the
{\em autopermutation group} of the code, a subgroup of $Aut(C)$ that restricts $g$ to qubit
permutations only. This restriction, however, is not necessary for the error models
and correction procedures considered here, and for the rest of the thesis, we consider $\R_C$ as a
equivalence relation over erasure patterns as defined in the previous paragraph, which
includes the full automorphism group.

If, for two correctable erasure patterns $\er{W},\er{K}$ we have 
$\er{W}\R_C\er{K}$, then failures during the erasure correcting procedure
of $\er{W}$ lead to error operators equivalent (under $\R_C$) to the error operators
resulting in failures during the erasure correcting procedure of $\er{K}$,
with the same probabilities.

Say we have $\er{W}$, and we measure an operator $\op{M}\in S$ in order
to attempt to correct a single erasure. Each
of possible $\mu$ failures during the measurement leads to the erasure patterns
$\er{W}_1,\cdots,\er{W}_\mu$ with respective probabilities 
$\Pr(\er{W}_1|\er{W}),\cdots,\Pr(\er{W}_\mu|\er{W})$. In order to correct $\er{K}$, 
we can apply a code automorphism 
to map it to $\er{W}$, then measure the
same operator $\op{M}$, which leads to the same possible erasure patterns
under failure, with exactly the same probabilities.
Applying the inverse of the automorphism leads to equivalent erasure patterns
under $\R_C$, with the same probabilities. Note that the application of these
automorphism does not need to correspond to operations in the physical
system. Permutations simply relabel the qubits, while applications of
$\Had$ and $\Pha$ relabel the stabilizers, so that one may simply
use this re-labeling to determine which operators need to be measured.

Thus, when analyzing the transition probabilities, we need only
consider transitions between equivalence classes under $\R_C$, greatly
reducing the number of transitions and erasure patterns that need to
be analyzed individually.

In practice, these equivalence classes can be found through 
computer algebra packages such as GAP \cite{GAP4} 
or MAGMA \cite{MAGMA}.
In the case of the Steane code, we have already seen some transversal (qubitwise)
unitary
operations that preserve the stabilizer (and thus the code space) -- namely
$\enc{\Had}$, $\enc{\Pha}^{\dagger}$. A quick
search through GAP demonstrates that the qubit permutations\footnote{
In this notation, a permutation $(a,b,c)$ means that qubit
 $a$ takes the place of $b$, $b$ takes the place of $c$, and $c$ takes
the place of $a$. Concatenation of two of these permutations
is equivalent to applying the rightmost permutation first, and
applying the permutations, in order, from right to left, just
like products of operators.}
\begin{subequations}
\begin{gather}
(1,2)(5,6)\\ 
(2,4)(3,5)\\ 
(2,3)(4,6,5,7)\\ 
(4,5)(6,7)\\
(4,6)(5,7),
\end{gather}
\end{subequations}
generate all permutations that preserve the code space. The group
generated by all these permutations and the unitaries described above
 is the automorphism group of the Steane code. It has order 168,
and it is isomorphic to the group of invertible $3\times 3$ matrices
in $GF(2)$.

\subsection{Z Measurement Threshold Calculation}
Focusing our attention on the error
model outlined in Section \ref{sec:ideal-error-model}, and on the 
Markov chain method outline in the previous section, we can calculate 
the recursion relation for the unintentional $\Z$ measurements
that can occur during teleportation. 

It has already been shown
that there is a code that has a $\Z$ measurement threshold of
$0.5$ for this error model \cite{klm-thr:2000}, but that code 
can only correct single $\Z$ measurements in a block. 
In Chapter \ref{ch:candidates} an alternative method for
correcting $\Z$ measurements was described, with the
advantage that only $\Z$ measurements are introduced as a result 
of failure, thus we can 
consider only the subgroup of $Aut(C)$ corresponding to the qubit
permutations, which is the code automorphism of the classical Simplex
code $[7,3,4]$ used to construct the Steane code. Any two
weight one erasure patterns are equivalent under the automorphism
relation.  Similarly, any two weight two erasure patterns are
equivalent as well.  This is not the case for weight three erasure
patterns, however. We know that $\enc{\Z}$ is a weight three Pauli
operator consisting of only $\Z$s and identities, so it will be an
element of the convex sum
$\I\ox\I\ox\I\ox\I\ox\er{Z}\ox\er{Z}\ox\er{Z}$, and thus this erasure pattern
 will 
be uncorrectable. By applying the permutations of the code automorphisms, we find that
only a subset of the weight three Pauli operators are generated. That is,
the equivalence class which includes this pattern does not include
all the possible erasure patterns, it only includes
\begin{equation}
\begin{array}{ccccccccccccl}
\I&\ox&\I&\ox&\I&\ox&\I&\ox&\er{Z}&\ox&\er{Z}&\ox&\er{Z}\\
\I&\ox&\I&\ox&\er{Z}&\ox&\er{Z}&\ox&\I&\ox&\I&\ox&\er{Z}\\
\I&\ox&\er{Z}&\ox&\I&\ox&\er{Z}&\ox&\I&\ox&\er{Z}&\ox&\I\\
\er{Z}&\ox&\er{Z}&\ox&\I&\ox&\I&\ox&\I&\ox&\I&\ox&\er{Z}\\
\I&\ox&\er{Z}&\ox&\er{Z}&\ox&\I&\ox&\I&\ox&\er{Z}&\ox&\I\\
\er{Z}&\ox&\I&\ox&\er{Z}&\ox&\I&\ox&\I&\ox&\er{Z}&\ox&\I\\
\er{Z}&\ox&\I&\ox&\I&\ox&\er{Z}&\ox&\er{Z}&\ox&\I&\ox&\I.
\end{array}
\end{equation}
All $28={7 \choose 3}-7$ remaining weight three erasure patterns are equivalent to each other. In order to
see that they are correctable, consider
\begin{equation}
\I\ox\I\ox\I\ox\er{Z}\ox\I\ox\er{Z}\ox\er{Z}
\end{equation}
We can correct the rightmost erasure by measuring the stabilizer
$\op{M}_3$ from Section \ref{sec:steane}, reducing the pattern to a
weight two erasure which is always correctable.

\begin{figure}
\includegraphics{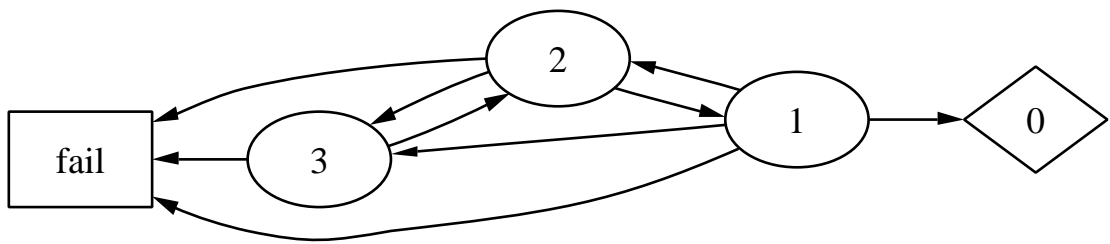}
\caption[Graph describing $\Z$ erasure correction.]{Graph describing the $\Z$ erasure correction procedure for the
  Steane code under the ideal error model. The probabilities
  associated with each transition are omitted for
  readability.\label{fig:z-graph}}
\end{figure}

Thus, we are left with four equivalence classes corresponding to
correctable erasure patterns of weight 0, 1, 2 and 3, and one
equivalence class corresponding to uncorrectable erasure patterns
(with weight 3 or more). The
transitions between the different equivalence classes can be
determined by analyzing a single erasure pattern in each class under
one erasure correcting step.  It suffices to say that for each case a
weight 4 stabilizer needs to be measured, which requires four $\CNOT$s
between the data and an ancilla, and the measurement of the
ancilla. Failures at any of these $\CNOT$s must be considered, which
is a simple but tedious process, described in detail in Appendix
\ref{ch:calc-z}, leading to the simplified pictorial view of Figure
\ref{fig:z-graph}.

\begin{figure}
\centering
\includegraphics[scale=0.5,angle=270]{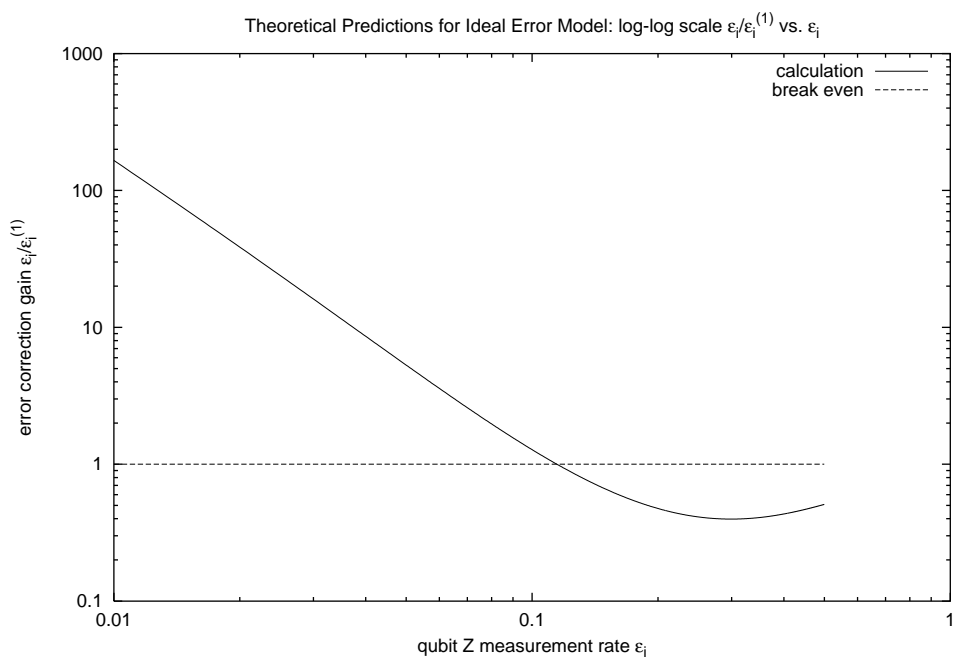}
\caption[Plot of $\frac{\e_i}{\e_i^{(1)}}$ under the ideal error model.]{Plot of $\frac{\e_i}{\e_i^{(1)}}$ under the ideal error model.
If the ratio is greater than $1$, the encoded measurement probability
is lower than the uncoded measurement probability.}
\end{figure}

From the Markov chain description one obtains the error rate recursion
relation
\begin{equation}\label{eqn:z-rec-rel}
\e_i^{(1)}=56\e_i^3+406\e_i^4+3878\e_i^5-129675\e_i^6+1164815\e_i^7+\cdots
\end{equation}
with a threshold of approximately $\e_i=0.115$.

The correctness of the leading term in \eqref{eqn:z-rec-rel} can be
quickly checked. First of all, because we are dealing with a distance
3 code, all measurement patterns of weight up to two can be corrected. That
immediately sets the exponent of the leading term to three.  In order
to determine the coefficient, we must count how many ways there are to
reach an uncorrectable measurement pattern with 3 errors during the correction
procedure, assuming of course we are doing a minimal number of
correcting rounds to be able to correct a weight three measurement
pattern.  First, we have seen that $\frac{7}{35}=\frac{1}{5}$ of the weight three
patterns are uncorrectable, that leaves the coefficient at
$\frac{1}{5}{7 \choose 3}=7$. Now consider the correction of a weight
two pattern: any single additional failure brings it back to a weight
three pattern, of which $\frac{1}{3}$ are uncorrectable, so we add
${7 \choose 2}\frac{1}{3}{3 \choose 1}=21$. In the case of weight one patterns, which
are also always correctable, we could have two failures at one, or one failure
at a time, adding ${7 \choose 1}\frac{1}{3}\left({3 \choose 2}+{3\choose 1}{3\choose 1}\right)=28$.
The total is, as expected, $56$.

\subsection{Full Erasure Threshold Calculation\label{subsec:full-thr-calc}}
In the lossy error model, things become a little more complicated.  In
particular, we have a greater variety of erasure patterns, which in
our case lead to a greater variety of equivalence classes and
transitions. In order to make it clear, note that the qubitwise
operations that are allowed in the definition of the automorphism of a
code are simply single qubit Clifford operations, which in essence
permute the labeling between the non-trivial Pauli matrices $\X,\Y,\Z$.
Thus, $\er{E}$ is invariant under these operations if we consider
a basis change given by the single qubit $\C_2$ operations, 
but $\er{X}, \er{Z}, \er{Y}$ are not. In fact, they
undergo the same label permutations as the Pauli matrices themselves.
That means that now, instead of keeping track of the weight of the
erasure patterns, we must keep track of the number of different types
of erasures. Some general properties still hold, such as
being able to correct any weight one or two erasure patterns,
and $\frac{4}{5}$ of all weight three patterns, as long as they consist 
of a single type of partial erasure. This is not as bad as it appears at first,
we can think of erasure patterns of a single type of partial erasure
acting together to result in an erasure pattern with different types of 
erasures. That is, we may consider
\begin{equation}\label{eqn:full-sample}
\er{E}\ox\I\ox\I\ox\er{Z}\ox\er{X}\ox\I\ox\I
\end{equation}
as a series of erasure patterns
\begin{subequations}\label{eqn:decomposition}
\begin{gather}
\er{X}\ox\I\ox\I\ox\I\ox\er{X}\ox\I\ox\I\\
\er{Z}\ox\I\ox\I\ox\er{Z}\ox\I\ox\I\ox\I
\end{gather}
\end{subequations}
applied to the same code block. So, because any weight two partial erasure
is correctable with the Steane code, any permutation of 
\eqref{eqn:full-sample} is correctable, even though it has weight three, so
in general, we need to keep track of a tuple that has the number of
$\er{E},\er{X},\er{Z}$ individually, and equivalence classes between
these tuples can be developed.

To simplify matters, because we are dealing with a self-orthogonal
CSS code, we can restrict ourselves with the measurement of $\Z$
stabilizers only through the application of $\CSIGN$s, so the only
types of errors that are introduced according to the lossy model
are either $\er{E}$ or $\er{Z}$. Following a decomposition
like \eqref{eqn:decomposition}, we find that the $\Z$ partial erasure
pattern will have a larger weight than the $\X$ partial erasure pattern
if any $\er{E}$ occurred in the original pattern. In that case,
because $\er{Z}$s are just as frequent as $\er{E}$s during computation in our model, 
it is wiser to correct a $\er{Z}$ first. If we choose to
correct a partial $\er{Z}$ without a corresponding partial $\er{X}$, 
and if we choose to correct an $\er{X}$ first when there are only
$\er{E}$, we ensure that the pattern consists of only $\er{E}$ and
$\er{Z}$ at all times. The analysis can then be restricted
to 2-tuples describing the number of full erasures and the
number of $\Z$ erasures -- graphically, they are described
by $[m,n]$ where $m$ is the number of $\er{E}$s, and
$n$ is the number of $\er{Z}$s, with the advantage that
the weight of the patterns is given by $m+n$.

The same correctability results in the previous section apply here, so we find that
all patterns of weight one and two are correctable, and that
for any equivalence class of patterns with weight three, $\frac{4}{5}$
of the patterns are correctable, and $\frac{1}{5}$ are not.
Again, we bundle all uncorrectable patterns into a single
class. The resulting graph can be seen in Figure \ref{fig:full-graph},
and the transition probabilities are described in detail in
Appendix \ref{ch:calc-full}. The main difference between the lossy
model and the ideal model is that for the lossy model we
consider measurement failures, with a probability of
failure $\dt$, along with a probability of $\CSIGN$ failure
$\e_l$ as usual. The transition probabilities are
given as a function of both considering them to be independent parameters\footnote{
By independence, I mean algebraic independence, not statistical.},
but in reality they are not. However, this dependence can be recovered
by writing one parameter in terms of the other in the general
expression obtained from the Markov chain description.

\begin{figure}
\includegraphics{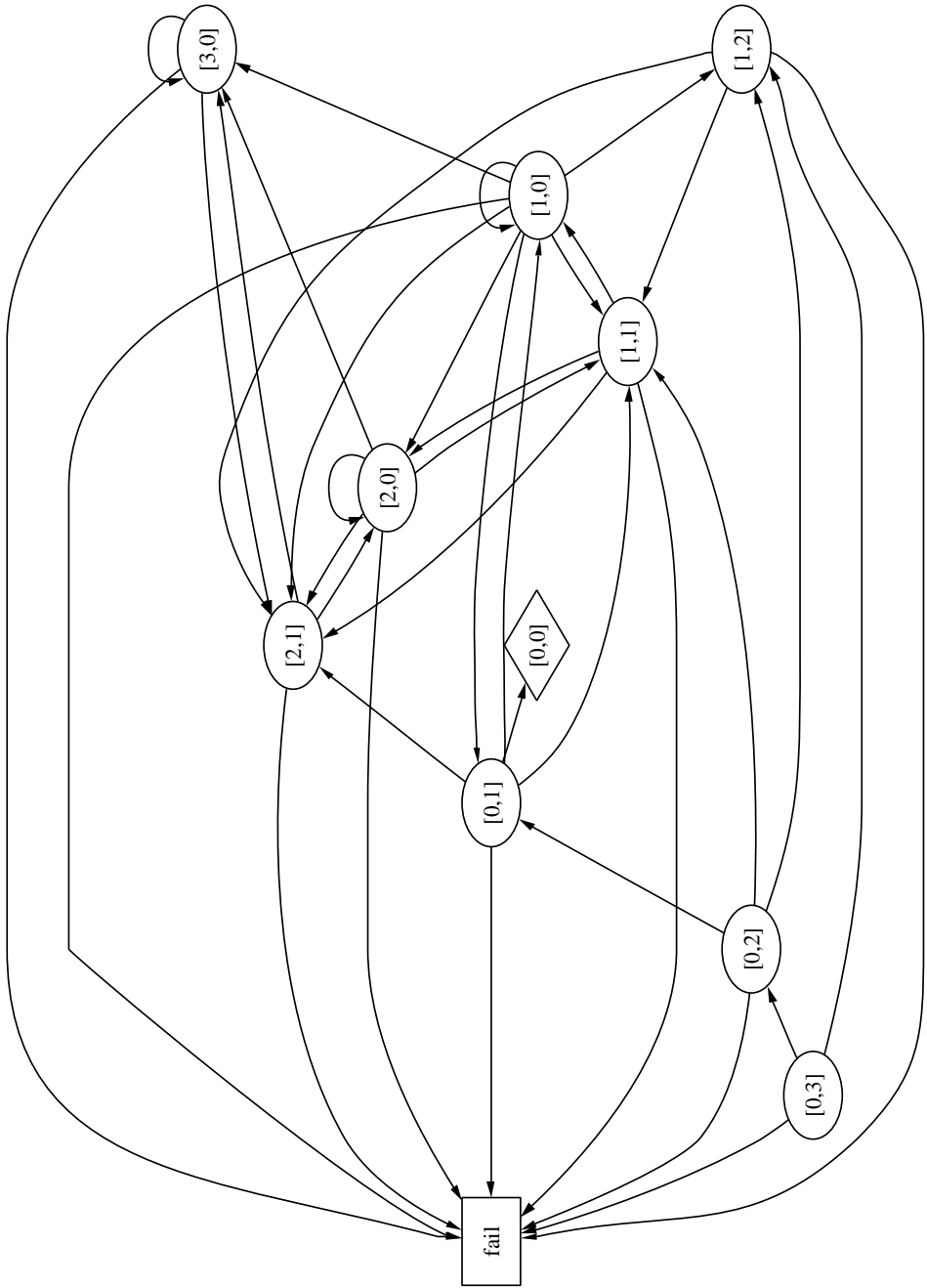}
\caption[Graph describing the full erasure correction procedure.]{Graph describing the erasure correction procedure for the
  Steane code under the lossy model. Each states is labeled $[m,n]$,
  where $m$ is the number of full erasures, and $n$ is the number of
  $\Z$ erasures. The probabilities associated  with each transition
  are omitted for readability.\label{fig:full-graph}}
\end{figure}

For the case where we take $\dt=0$
\begin{equation}
\e_l^{(1)}=350\e_l^3+4739\e_l^4-12404\e_l^5-355600\e_l^6-3087110\e_l^7+\cdots=\frac{1}{2}\e_l
\end{equation}
yielding a threshold of approximately $\e_l=0.0324$.

For the case where we take $\dt=\e_l$
\begin{equation}
\e_l^{(1)}=1050\e_l^3+33173\e_l^4-46242\e_l^5-6861701\e_l^6-118743847\e_l^7+\cdots=\frac{1}{2}\e_l
\end{equation}
which yields a threshold of approximately $\e_l=0.0178$. This threshold is
only valid if the encoded qubit measurement failure rate $\dt^{(L)}$ also
vanishes for $L\to\infty$. Because of the structure of the CSS code,
where the encoded states are superposition of classical codewords,
measurement failures can be corrected for. In the case of the Steane
code, the classical codewords that are measured are codewords in
the classical $[7,4,3]$ Hamming code, so one can correct at least up to $2$
erasures, which are the consequence of measurement failures. Thus, the
encoded failure rate for measurements in the first level of encoding is
\begin{equation}
\dt^{(1)}=\sum_{i=3}^7{7 \choose i}\dt^i(1-\dt)^{7-i},
\end{equation}
assuming that classical processing needed for classical error correction
is perfect. This yields a threshold of approximately $\dt=0.25$, which guarantees the
validity of both $\e_l$ thresholds described above.

These two cases, $\dt=0$ and $\dt=\e_l$, are the extremal cases for the
calculation of the threshold. In the limit where loss in the $\CSIGN$s
is much more likely than at the detectors, we have $\dt=0$, while in the
limit where loss at the detectors is of the same order as loss in the
$\CSIGN$s, we have $\dt=\e_l$. Because of the general construction of the $\CSIGN$
based on photon mode teleportation, there will be more than one detector involved,
so the probability that there is a failure in the $\CSIGN$ teleportation will always
be higher than the probability of a single detector failing, justifying this
choice of upper and lower bounds.

\begin{figure}
\centering
\includegraphics[scale=0.8, angle=270]{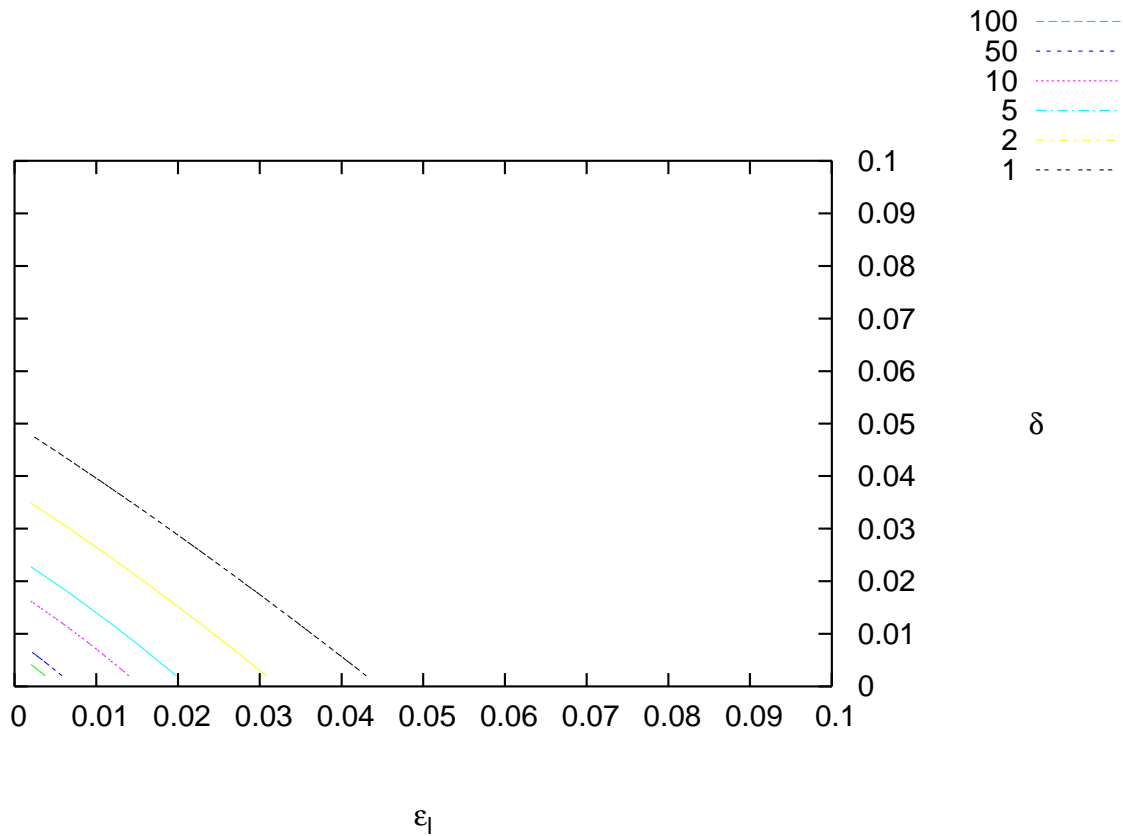}
\caption[Encoded error rate as function of gate and detector failures.]{Contour plot of $\frac{\e_l}{\e_l^{(1)}}$ as a function of $\e_l$ and $\dt$ under
the lossy error model. The threshold is given by $\frac{\e_l}{\e_l^{(1)}}=\frac{1}{2}$, and the
greater this ratio is, the greater the gain for using erasure correction codes.}
\end{figure}

\begin{figure}
\centering
\includegraphics[scale=0.5,angle=270]{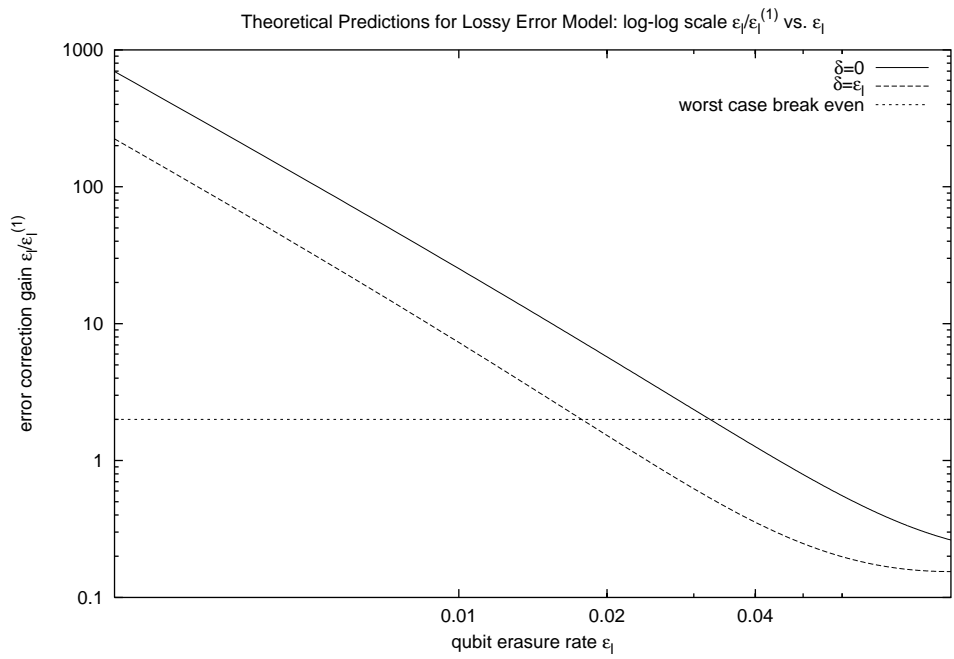}
\caption[Plots of $\frac{\e_l}{\e_l^{(1)}}$ under the lossy error model, $\dt=0$ and $\dt=\e_l$.]{Plots of $\frac{\e_l}{\e_l^{(1)}}$ under the lossy error model, for
extremal probabilities of detector failure, $\dt=0$ and $\dt=\e_l$.}
\end{figure}

%% file: Simulation.tex
\chapter{Simulation}
\markright{Simulation}

In order to verify the theoretical predictions about the recursion relation
for the Steane code, a Monte Carlo simulation of the erasure correcting
procedure was made following the principles previously used by 
Zalka \cite{zalka:1996},
which I review here for completeness. The results of the simulations for
the ideal and lossy error models are presented and contrasted with
the theoretical predictions and calculations of Chapter \ref{ch:erasure-threshold}.

\section{Background}
Fully simulating the states of the quantum computer, aside from having
an exponentially high cost in resources, is superfluous for the 
determination of the performance of an error correcting code. Since we
are only concerned with whether the error correction procedure is
able to eliminate all erasures, we can keep track of only that
aspect of the computation. This is the same basic idea used by
Zalka in \cite{zalka:1996}, but applying various simplifications
due to the simpler handling of erasures as well as new insights
into fault-tolerant universal computation by various authors
\cite{gottesman-thesis:1997,gottesman-chuang:1999,zlc:2000}.

While in the case of general errors one must keep track of how computation
and error correction propagates error operators that remain undetected, 
when dealing with erasures
we simply need to keep track of where they appear during computation
and erasure correction. The error operators that are kept track of
are simply Pauli operators over the entire code block.

It is crucial to note that this setup allows only for the simulation
of gates that map the Pauli group onto itself under conjugation,
since, by assumption, all the erasure operators considered here are
described by convex sums of Pauli operators -- the Clifford group
$\C_2$ described previously. This is not a universal set for quantum
computation, but it is enough for error correction.
In order to make a universal set, one can include,
for example, gates conditional on measurement outcomes.  This was the
method employed by Shor in the seminal paper on fault-tolerant
computation to demonstrate the construction of an encoded Toffoli for
the Steane code \cite{shor:1996}, and it was further generalized to
demonstrate the construction of other encoded gates such as the
$\frac{\pi}{8}$ gate and the controlled phase gate
\cite{gottesman-chuang:1999,zlc:2000}.

Due to the fact that the encoded Toffoli can be built out of  encoded
Clifford gates, it is intuitive that a threshold for Clifford gates
would also be a threshold for the Toffoli. This was analyzed and
quantified carefully by Gottesman \cite{gottesman-thesis:1997} for the
Steane code, where it was demonstrated that indeed the threshold for
universal computation with Clifford gates plus the Toffoli is only
$\sim 5\%$ lower than the threshold for  Clifford gates. In the case
of linear optics, the threshold would depend on the detector failure
rate, since in order to construct these fault-tolerant gates necessary
for universal computation measurements and operations conditioned on
measurements are necessary. However, we have shown in Section \ref{subsec:full-thr-calc} that
the operation
of the Clifford gates themselves, and the measurement of the syndromes
already places enough constraint in the detector efficiency so that
a result similar to the $5\%$ approximation of the universal
threshold given by Gottesman should still hold.

\section{Data Structures and Algorithms}
\begin{figure}
\centering
\includegraphics{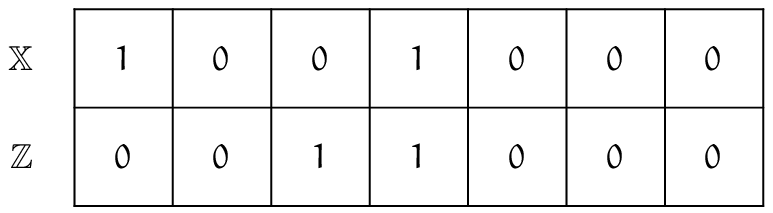}
\caption[Internal representation of 
a stabilizer operator
in the Monte Carlo simulation.]{Internal representation of $\er{X}\ox\I\ox\er{Z}\ox\er{E}\ox\I\ox\I\ox\I$ in the Monte Carlo simulation.\label{fig:array}}
\end{figure}

The simulation consists, in essence, of two arrays of bits of length
7, each position representing one of the 7 qubits that constitute a
codeword in the Steane code. One of the arrays, called the $\er{X}$
array, indicates the application of an $\X$ erasure on each qubit,
while the other array, called the $\er{Z}$ array, indicates the
application of a $\Z$ erasure on each qubit -- this is depicted in
Figure \ref{fig:array}, and can be thought of as the
side information mentioned in Section \ref{sec:erasure-channel}. 
One can represent full erasures as well, as
$\er{X}$ and $\er{Z}$ commute, and
\begin{subequations}\label{eqn:erasure-rules}
\begin{eqnarray}
\er{Z}\left(\er{X}\left(\rho\right)\right)&=&\frac{1}{2}\left(\er{X}(\rho)+\Z\er{X}(\rho)\Z\right)\\
&=&\frac{1}{4}\left(\rho+\X\rho\X+\Z\rho\Z+\Y\rho\Y\right)\\
&=&\er{X}\left(\er{Z}\left(\rho\right)\right)\\
&=&\er{E}\left(\rho\right).
\end{eqnarray}
\end{subequations}
In the simulation, a full erasure in qubit $i$ is therefore
represented by setting the $i$th bit in both the $\er{X}$ and the
$\er{Z}$ arrays.

As has been stated, the indication of any type of erasure over some
qubit is by no means an indication that an error has indeed
happened. The convex sum representation of the erasure superoperator
can be interpreted, instead, as an indication that an error {\em might}
have happened at that location. It is still necessary to measure the
syndrome of the qubit erased to determine which of the possible Pauli
operators has occurred. As described in Section \ref{sec:error-model},
the only sources of erasure that are considered here are the failures
in teleportation of the $\CSIGN$, and photon loss in the teleportation
and detection of qubits. It has also been shown that these are all
erasures, so it is always known when the data is corrupted by these
failures, 
and one may simply try
to generate a large number of ancillae in the desired state and
discard all the ones that contain erasures.

In the simulation, we assume the erasure correction procedure is
applied after a single application of an encoded $\CSIGN$
gate resulting in erasures. Because the $\CNOT$ and the Hadamard are transverse on the
Steane code (see  Section \ref{subsec:steane-ft-gate}), so is the $\CSIGN$, and it
realizes the encoded $\CSIGN$ itself. In that case, we assume each of
the gates fails independently at the target and at the control. Under
the ideal error model, the only type of failure that may occur is a
$\Z$ measurement, and it is localized to either the control or the target
independently. Under the lossy model, however, a teleportation failure at the $\CSIGN$
induces a full erasure locally and a $\Z$ erasure at the gate
counterpart -- for example, a failure at the target causes a full
erasure at the target and a $\Z$ erasure at the control.  Thus, the
error pattern on a code block depends not only on the failures that
occurred locally, but also on the failures that occurred on a block with
which it was interacting, but again, because of the rule given by
\eqref{eqn:erasure-rules}, this is easy to compute. Under both erasure
models one simulates the initial state of the block by flipping a
biased coin for each qubit to determine whether it should have an
erasure or not, and  in the case of the lossy error model, there are
two passes: one determines  whether there should be a full erasure,
and the other determines whether it should have a $\Z$ erasure if it
did not have a full erasure.

Once the erasure pattern has been generated, multiple rounds of single
erasure correction are attempted until either the code block is
erasure free, or the erasure pattern is uncorrectable. Since each
single erasure correction step requires the measurements of a weight
four stabilizer operator (in the case of the Steane code), any of
these gates may fail at either the control or the target side,
independently, so once again biased coins are flipped.  In this case,
the difference between the two error models is a little more
salient. For the ideal model, we can simply count how many failures
there were on the ancilla side of the stabilizer measurement in order
to determine if the error correction will succeed or fail, and then
determine which gates failed on the data side so that the erasure
pattern may be updated for the next correction attempt.  Under the
lossy model, however, one must keep track of which $\CSIGN$s with the
ancilla failed, because those failures will induce erasures on the
data side, which must be remembered for the next correction attempt,
as well as keeping track of which failures occurred on the data
side. Which qubits will need to interact with the ancilla is
determined by the erasure pattern, as described in Sections \ref{subsec:steane-er-corr}
and \ref{subsec:z-measurement}.

The lossy model has another distinction in that we allow for erasures
to occur at the measurement of the ancilla, although at a rate $\dt$
different from the gate failure rate. In reality $\e_l$ and $\dt$
are not independent, since constructions of the $\CSIGN$ gate itself
depend on the measurement of ancillary photonic modes. The exact
relationship is highly dependent on which $\CSIGN$ construction is
being used, and since each has its own advantages, it was decided to
make the simulation take $\e_l$ and $\dt$ as independent parameters that
can be adjusted as desired.

For each gate failure rate the simulation is run until at least one
thousand blocks are found to be uncorrectable (for the low
probabilities of failure), or until at least a large number of blocks,
proportional to $\frac{1}{\e}$, have been simulated (for high
probabilities of failure). Statistics, such as how many blocks were
simulated, how many were uncorrectable, etc. are output into a file,
and the process is repeated for a different random number generator
seed. After 20 such iterations, the individual statistics are
compiled to obtain a mean and a standard deviation for the encoded
failure rate at every gate failure rate (and detector failure rate).

\section{Results}
\subsection{Ideal error model}
\begin{figure}
\centering
\includegraphics[scale=0.5,angle=270]{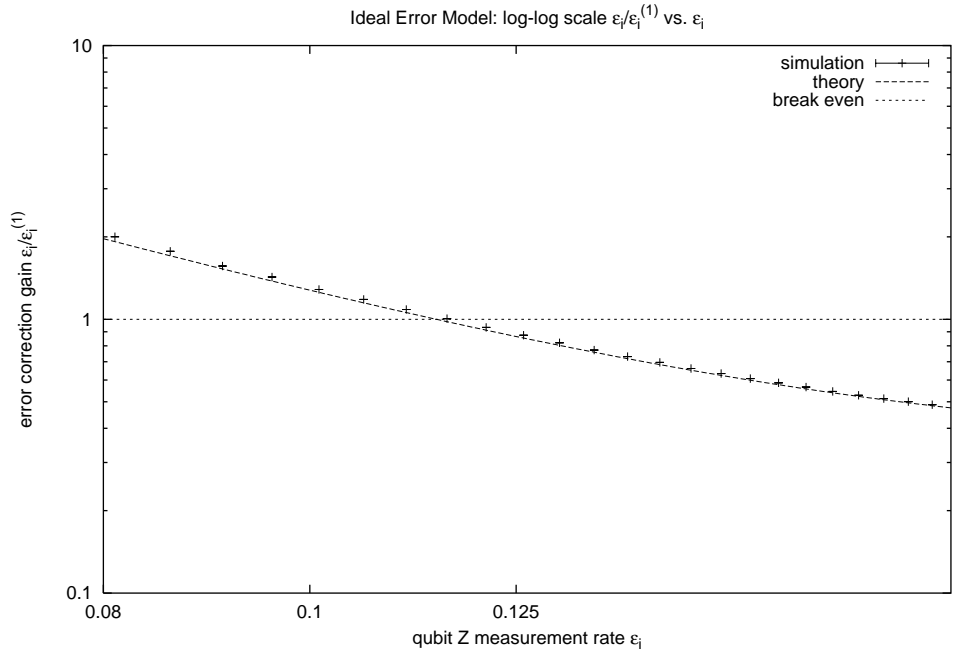}
\caption[Probability of failure for teleportation failure rate 
$\e$.]{Probability of failure for teleportation failure rate 
$\e$ using 6 erasure correction attempts.\label{fig:z-results}}
\end{figure}

As can be observed in Figure \ref{fig:z-results}, the simulation
matches the predicted values very closely.
Because of repeated trials with different seeds for the random number
generators, the error bars for the simulation results are very small,
and we can take the variance in the results to be insignificant.

\subsection{Lossy error model}

\begin{figure}
\centering
\includegraphics[scale=0.5,angle=270]{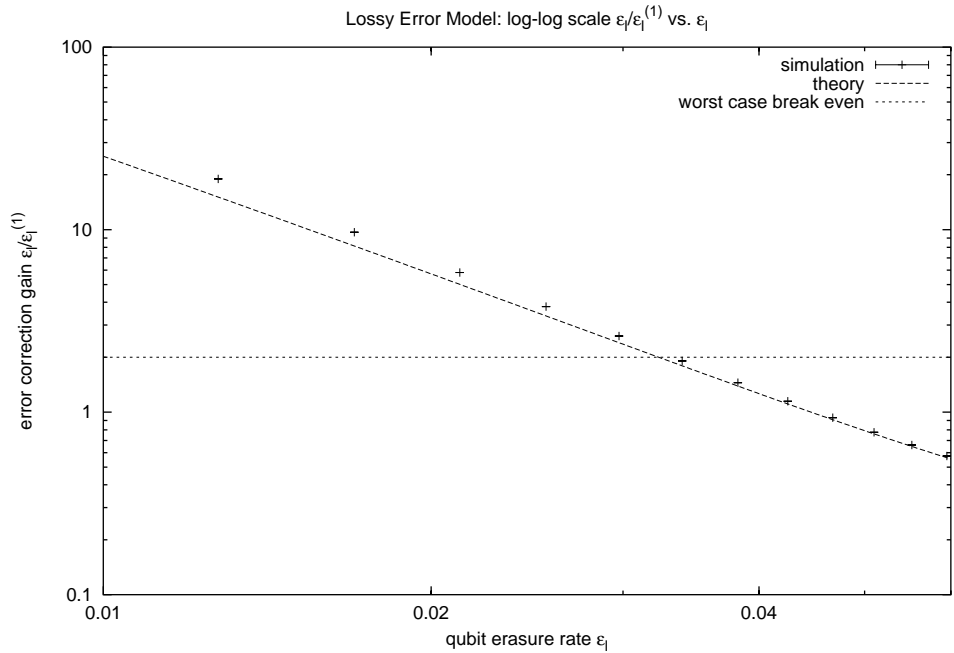}
\caption[Probability of failure given $\dt=0$.]{Probability of failure given $\dt=0$ with 20 
erasure correction attempts.\label{fig:full-results-d=0}}
\end{figure}

\begin{figure}
\centering
\includegraphics[scale=0.5,angle=270]{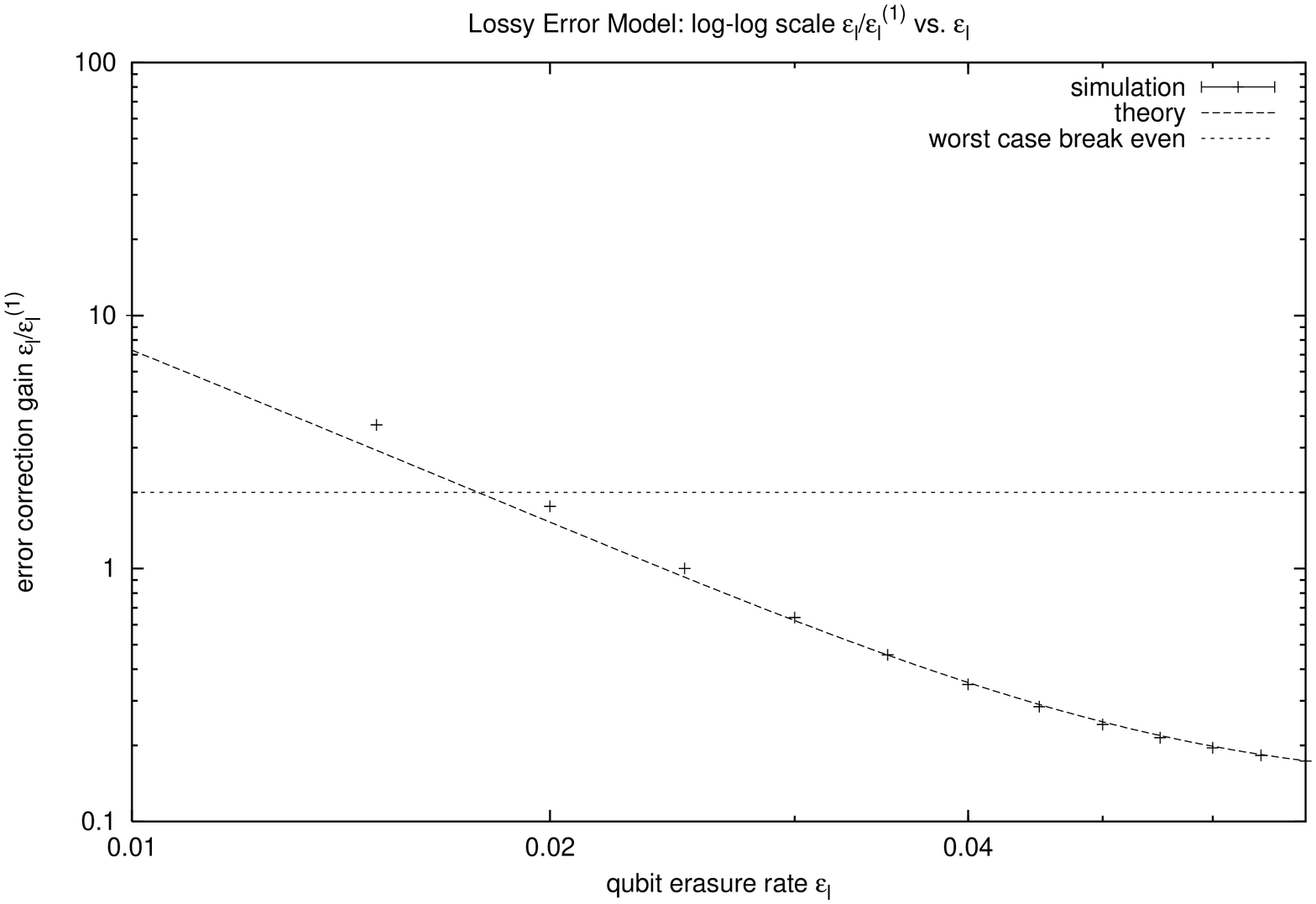}
\caption[Probability of failure given $\dt=\e_l$.]{Probability of failure given $\dt=\e_l$ with 20 
erasure correction attempts.\label{fig:full-results-d=e}}
\end{figure}

Again, as can be seen in Figures \ref{fig:full-results-d=0} and 
\ref{fig:full-results-d=e}
the error bars in the simulation are so small as to be
insignificant, but in this case there is a discrepancy between the
simulation results and the calculated values for probabilities
below the threshold. 

There are a number of possible reasons for this behavior, but the main
source of error appears to be the recursion relation itself and using
computers to numerically approximate its value at different gate
failure probabilities.  While it is necessary to perform the 20
correction attempts in order to have the same precision in the leading
terms of  the recursion relation as in the ideal case, that makes the
recursion relation extremely long, with very large integer
coefficients, both positive and negative. By construction, this
recursion relation must stay bounded to values between $0$ and
$1$. However, while the mathematical packages have infinite precision
for integer handling, the same is not true for floating point
calculations, necessary in the evaluation of these functions. The
finite precision of these calculations used to generate these plots
should potentially be the main source of error.
The recursion relations themselves are correct.

A different approach to verify the correctness of the
recursion relations would be to fit polynomial
curves to the simulation results. However, in order to
do so reliably, one needs to run simulations at extremely
low probabilities, increasing the error on the curves significantly
simply because not enough statistical sampling is done.
This is the same problem observed by Steane \cite{steane:2003},
and it is a common problem of Monte Carlo simulations
at low event probabilities.

%% file: Conclusion.tex
\chapter{Conclusion}
\markright{Conclusion}

Through theoretical calculations and verification by simulation, we have
found that in an error model consisting only of ideal teleportation
failures, there is a threshold of $\e_i\approx0.115=11.5\%$, and that for an ideal
model consisting of photon loss and replacement during teleportation,
the threshold is at least $\e_l\approx0.0178=1.78\%$ as long as the detector failure rate at most
equal to the gate failure rate. A mixed error model consisting
of both types of failure would have a threshold somewhere in between
these two as long as the total probability of failure $\e$ is bounded by these two values, that is
\begin{equation}
\e_l<\e<\e_i.
\end{equation}
In the limit where the photon loss dominates, one expects the threshold to 
be closer to $\e_l$, while in the case where ideal failures
dominate, the threshold would be higher and closer to $\e_i$.

%
The results presented here need to be considered carefully. 
The correction procedure used for the threshold calculation can correct either a single
general error at an unknown location, or two (and at most three) erasure patterns.
It cannot handle a mixed error model involving both errors and erasures. For
realistic quantum computation we need to consider all of these sources of error,
so one would need to consider a modified
correction procedure.

However, the main result to be taken from this work is that the recursion
relation for the erasure correction procedure can be derived exactly, and this
may lead to new techniques of analyzing the performance of erasure correcting
codes, or even general error correcting codes. 
The erasure correcting procedure used here is not fully optimized, 
since it was tailored to simplify the exact analysis, but there are
many avenues that may be taken to improve these bounds on the erasure threshold,
including different syndrome measurement 
techniques \cite{steane-filter:2002,steane-css-networks:2003}.
It is important to note that the thresholds presented here are well
above the general error thresholds presented in \cite{steane:2003,zalka:1996},
and although the threshold for the ideal case if substantially worse
than the $\e_{klm}=0.5$ threshold  given in \cite{klm-thr:2000},
my result considers a more general and more realistic error model.

\section{Future Work}

In order to make the results presented here of direct applicability
to linear optics quantum computing, one will have to take the different
construction of quantum gates into account, and determine how $\e_l$ depends
on the detector efficiency $\dt$. There has been some previous work in that
area \cite{ralph:2003}, 
but none that I am aware relating detector efficiency directly to 
a threshold for quantum computing. This relationship would be of extreme
importance in determining what the requirements are for the performance of
single photon detectors. The results presented here are the first step
in that direction.

As mentioned before, one would need to consider larger codes or different
error correction procedures in order to be able
to correct both erasures and general errors. Recently a large survey
of the performance of different CSS codes has been published \cite{steane:2003},
and it has even been pointed out how the $[[23,1,7]]$ Golay code provides
a better threshold for general errors than the Steane code. A modification
of the simulation used to obtain the results presented here should be straight
forward, especially with the insights into how to perform erasure correction
by using information about the automorphism of the code. There is no
comprehensive study of how different codes compare for an erasure error
model, although there have been many papers dealing with erasure
correction codes \cite{grass-etal:1997,grassl-bch:1999}, 
so a survey of erasure thresholds along the lines
of \cite{steane:2003} could be elucidating.

Although for the error model presented here it would be unreasonable to
choose anything but a CSS code, for other error models that may have
erasures at any gate the same may very well not be true. A modification
of the simulation could be done based on ideas on how to encode
stabilizer codes in binary string (presented in 
\cite{calderbank-etal:1998}), but how to
modify the code to make it general and flexible enough for different
error models is more involved. A natural first step would be
to simply use general error models for gate and memory error so
comparisons can be made between known stabilizer codes and the
CSS codes presented in \cite{steane:2003}.

Finally, the most interesting aspect of this research, in my opinion, has been the
calculation of the recursion relations by describing the erasure
correction procedure as a Markov chain. An investigation into how to extend
this work into the general error model could yield precise insights
into what governs the relationship between different code
parameters, such as the weight distributions, the automorphisms of the
code, etc. and the thresholds obtained with that code.

%% file: calc-z.tex
\chapter{Calculation for $\Z$ measurement Threshold\label{ch:calc-z}}
\markright{Calculation for $\Z$ measurement Threshold}

Following the procedure described in Section \ref{sec:era-corr},
we take $\e_i$ to be the probability of a $\Z$ measurement occurring due
to teleportation failure.

The error free pattern, that is, a measurement pattern of weight $0$, 
is an absorbing state, so the
only non-zero probability of transition is into itself -- nothing needs to
be corrected. The same
is true for the uncorrectable measurement patterns, represented by the
{\em fail} state.
\begin{eqnarray}
\Pr\left(i|0\right) &=& \delta_{i,0}\\
\Pr\left(i|\text{{\em fail}}\right) &=& 0\\
\Pr\left(\text{{\em fail}}|\text{{\em fail}}\right) &=& 1.
\end{eqnarray}
Regardless of what the current state, the probability of correcting a
single $\Z$ measurement is given by the probability that there are no failures 
{\em at all} is the correction procedure, that is, all three teleportations
succeed, or
\begin{equation}
\Pr(\text{{\em success}})=(1-\e_i)^3.
\end{equation}
The probability of staying in an measurement pattern of the same weight is
$0$ since any teleportation failure in the correction procedure prevents the
restoration of the previous measured qubit.

Many of the transition probabilities in the measurement correction procedure used
here are zero. Exactly which transition are possible given the procedure
follows from some simple rules. Given some $\Z$ measurement pattern of weight $i$
\begin{itemize}
\item only one measurement is corrected at a time, resulting in an measurement pattern
of weight $i-1$ in the best case. 
\item otherwise, the weight of the measurement pattern can only increase or stay
the same (Murphy's Law for $\Z$ measurement correction).
\end{itemize}
Using general cases described here, we calculate all the non-zero probabilities of transition,
omitting detailed description of how they were obtained for brevity. All probabilities
of transition equaling zero are omitted. 
\section{Transitions from weight 1 patterns}
\begin{subequations}
\begin{eqnarray}
\Pr(0|1)&=&\Pr(\text{{\em success}})\\
\Pr(2|1)&=&{3 \choose 1}\e_i(1-\e_i)^2\\
\Pr(3|1)&=&\frac{2}{3}{3 \choose 2}\e_i^2(1-\e_i)\\
\Pr\left(\text{{\em fail}}|1\right)&=&\frac{1}{3}{3 \choose 2}\e_i^2(1-\e_i)+\e_i^3
\end{eqnarray}
\end{subequations}
\section{Transitions from weight 2 patterns}
\begin{subequations}
\begin{eqnarray}
\Pr(1|2)&=&\Pr(\text{{\em success}})\\
\Pr(3|2)&=&\frac{2}{3}{3 \choose 1}\e_i(1-\e_i)^2\\
\Pr\left(\text{{\em fail}}|2\right)&=&\frac{1}{3}{3 \choose 1}\e_i(1-\e_i)^2+{3 \choose 2}\e_i^2(1-\e_i)+\e_i^3
\end{eqnarray}
\end{subequations}
\section{Transitions from correctable weight 3 patterns}
\begin{subequations}
\begin{eqnarray}
\Pr(2|3)&=&\Pr(\text{{\em success}})\\
\Pr\left(\text{{\em fail}}|3\right)&=&{3 \choose 1}\e_i(1-\e_i)^2+{3 \choose 2}\e_i^2(1-\e_i)+\e_i^3
\end{eqnarray}
\end{subequations}
\section{Initial distribution}
Since we are only considering a single type of error -- namely, unintentional
$\Z$ measurements due to teleportation failure -- this calculation 
is quite simple, yielding
\begin{subequations}
\begin{eqnarray}
\Pr(j)&=&{7 \choose j}\e_i^j(1-\e_i)^{7-j}, j<3\\
\Pr(3)&=&\frac{4}{5}{7 \choose 3}\e_i^3(1-\e_i)^{4}\\
\Pr\left(\text{{\em fail}}\right)&=&\frac{1}{5}{7 \choose 3}\e_i^3(1-\e_i)^{4}+\sum_{j=4}^7{7 \choose j}\e_i^j(1-\e_i)^{7-j}
\end{eqnarray}
\end{subequations}

%% file: calc-full.tex
\chapter{Calculation for Full Erasure Threshold\label{ch:calc-full}}
\markright{Calculation for Full Erasure Threshold}

Following the procedure described in Section \ref{sec:era-corr},
we take $\e_l$ to be the probability of a full erasure
occurring, and $\dt$ to be the probability that a single detector
will fail. The general idea is to adapt the procedure used for correcting
$\Z$ measurements and use it to correct $\Z$ erasures, and use the standard
fault-tolerant stabilizer measurement of Section \ref{subsec:ft-stab-meas}
to correct $\X$ erasures.

The probability that there will be a failure in any of the detectors
in the measurement of the ancilla (in the fault-tolerant stabilizer
measurement) is
\begin{equation}
\Pr(\text{{\em ancilla det.}}) = 1 - (1-\dt)^4,
\end{equation}
since all $\Z$ stabilizer operators of the Steane code have weight $4$.
The error free state $[0,0]$ is an absorbing state, so the
only non-zero probability of transition is into itself. The same
is true for the uncorrectable errors, represented by the
{\em fail} state.
\begin{eqnarray}
\Pr\left([i,j]|[0,0]\right) &=& \delta_{i,0}\delta_{j,0}\\
\Pr\left([i,j]|\text{{\em fail}}\right) &=& 0\\
\Pr\left(\text{{\em fail}}\big|\text{{\em fail}}\right) &=& 1.
\end{eqnarray}

Regardless of what the current state is, the probability of correcting the
single erasure we were targeting is
given by the probability that there are no failures {\em at all}.
In the case of an $\X$ erasure, that means
\begin{equation}
\Pr(\text{{\em success}}|\er{X})=(1-\e_l)^8(1-\Pr(\text{{\em ancilla det.}})).
\end{equation}
In the case of a $\Z$ erasure, we need to consider the qubit teleportation
described in Figure \ref{fig:z-measurement-teleport}. There we have a single
entangling gate, the $90^\degg$ rotation about $\Z\ox\Y$, that like the $\CSIGN$ is implemented through mode
teleportation, so a photon is lost with the same probability $\e_l$. Then
each of the two qubits is measured with an independent probability of failure $\dt$.
Thus, the probability that a photon is lost in this teleportation is given by
\begin{equation}
\Pr(\text{{\em loss}}) = \e_l+(1-\e_l)\left[{2 \choose 1}\dt(1-\dt)+\dt^2\right].
\end{equation}
We also need to measure the $\Z$ erasure in order to apply the same correction
procedure as in the $\Z$ measurement, yielding
\begin{equation}
\Pr(\text{{\em success}}|\er{Z})=(1-\dt)\left(1-\Pr(\text{{\em loss}})\right)^3.
\end{equation}

When attempting to correct a $\er{Z}$, the probability of staying in
the same state is $0$, since any failure increases the weight of the
erasure pattern and re-randomizes the phase of the qubit with a $\Z$ erasure,
while failure during the initial measurement transforms a $\er{Z}$ into a
$\er{E}$.
On the other hand, when there are only full erasures and an $\er{X}$ must be corrected
(recall $\er{X}$ are always corrected first, by design), in order to
stay in the same erasure pattern, either only the measurement must fail,
the $\CSIGN$ on the target failed on the data side, or the $\CSIGN$ failed
{\em only} on the ancilla side
\begin{equation}
\Pr(\text{stay }\er{E}) = (1-\e_l)^6[(1-\e_l)^2\Pr(\text{{\em ancilla det.}})+\e_l+(1-\e_l)\e_l]
\end{equation}

Many of the transition probabilities in the erasure correcting procedure used
here are zero. Exactly which transitions occur with non-zero
probability, given the procedure,
follows from some simple rules. Given some erasure pattern described by
$[m,n]$ where $m$ is the number of full erasures and $n$ is the number of
$\Z$ erasures,
\begin{itemize}
\item the type of erasure ($\er{X}$ or $\er{Z}$) that is most abundant is corrected first.
\item only one erasure is corrected at a time, but when a full erasure
is corrected, it becomes a $\Z$ erasure, resulting in $[m-1,n+1]$ in the best case. 
If a $\Z$ erasure is being corrected, the result is $[m,n-1]$ in the best case.
\item a $\Z$ erasure may be transformed into a full erasure if, when the qubits is
measured in the $\Z$ basis, the detector fails.
\item failure during a $\Z$ erasure correction results in the $\Z$ erasure staying
as a $\Z$ erasure, but introducing full erasures in the other three qubits in the
subsystem. Thus, only the number of full erasures can increase due to failure during a $\Z$ erasure
correction -- by a minimum of 1 and a maximum of 3.
\item Failures during an $\X$ erasure correction results in 
the number of either types of erasures increasing
or stay the same.
\end{itemize}

Using the general cases described here, we calculate all the non-zero probabilities of transition,
omitting detailed description of how they were obtained for brevity. All probabilities
of transition equaling zero are omitted, and the probabilities of transition
to the {\em fail} state are chosen such that the probabilities of transition from
any given state sum up to $1$.
\section{Transitions from [0,1]}
\begin{subequations}
\begin{eqnarray}
\Pr([0,0]|[0,1]) &=& \Pr(\text{success}|\er{Z})\\
\Pr([1,0]|[0,1]) &=& \dt\\
\Pr([1,1]|[0,1]) &=& (1-\dt){3 \choose 1}\Pr(loss)[1-\Pr(loss)]^2\\
\Pr([2,1]|[0,1]) &=& (1-\dt)\frac{2}{3}{3 \choose 2}[\Pr(loss)]^2[1-\Pr(loss)]
\end{eqnarray}
\end{subequations}
\section{Transitions from [1,0]}
\begin{subequations}
\begin{eqnarray}
\Pr([0,1]|[1,0]) &=&\Pr(\text{success}|\er{E})\\
\Pr([1,0]|[1,0])&=&\Pr(\text{stay }\er{E})\\
\Pr([1,1]|[1,0])&=&(1-\e_l)^3{3 \choose 1}\e_l(1-\e_l)^2\\
\Pr([2,0]|[1,0])&=&{3 \choose 1}\e_l(1-\e_l)^2(1-\e_l)^2\\
\Pr([2,1]|[1,0])&=&\frac{2}{3}{3 \choose 1}\e_l(1-\e_l)^2{2 \choose 1}\e_l(1-\e_l)\\
\Pr([1,2]|[1,0])&=&\frac{2}{3}(1-\e_l)^3{3 \choose 2}\e_l^2(1-\e_l)\\
\Pr([3,0]|[1,0])&=&\frac{2}{3}{3 \choose 2}\e_l^2(1-\e_l)\\
\end{eqnarray}
\end{subequations}
\section{Transitions from [1,1]}
\begin{subequations}
\begin{eqnarray}
\Pr([1,0]|[1,1])&=&\Pr(\text{success}|\er{Z})\\
\Pr([2,0]|[1,1])&=&\dt\\
\Pr([2,1]|[1,1])&=&(1-\dt)\frac{2}{3}{3 \choose 1}\Pr(loss)[1-\Pr(loss)]^2
\end{eqnarray}
\end{subequations}
\section{Transitions from [2,0]}
\begin{subequations}
\begin{eqnarray}
\Pr([1,1]|[2,0])&=&\Pr(\text{success}|\er{E})\\
\Pr([2,0]|[2,0])&=&\Pr(\text{stay }\er{E})\\
\Pr([2,1]|[2,0])&=&\frac{2}{3}(1-\e_l)^3{3 \choose 1}\e_l(1-\e_l)^2\\
\Pr([3,0]|[2,0])&=&\frac{2}{3}{3 \choose 1}\e_l(1-\e_l)^2(1-\e_l)^2
\end{eqnarray}
\end{subequations}
\section{Transitions from [0,2]}
\begin{subequations}
\begin{eqnarray}
\Pr([0,1]|[0,2])&=&\Pr(\text{success}|\er{Z})\\
\Pr([1,1]|[0,2])&=&\dt\\
\Pr([1,2]|[0,2])&=&(1-\dt)\frac{2}{3}{3 \choose 1}\Pr(loss)[1-\Pr(loss)]^2
\end{eqnarray}
\end{subequations}
\section{Transitions from [3,0]}
\begin{subequations}
\begin{eqnarray}
\Pr([2,1]|[3,0])&=&\Pr(\text{success}|\er{E})\\
\Pr([3,0]|[3,0])&=&\Pr(\text{stay }\er{E})
\end{eqnarray}
\end{subequations}
\section{Transitions from [0,3]}
\begin{subequations}
\begin{eqnarray}
\Pr([0,2]|[0,3])&=&\Pr(\text{success}|\er{Z})\\
\Pr([1,2]|[0,3])&=&\dt
\end{eqnarray}
\end{subequations}
\section{Transitions from [2,1]}
\begin{subequations}
\begin{eqnarray}
\Pr([2,0]|[2,1])&=&\Pr(\text{success}|\er{Z})\\
\Pr([3,0]|[2,1])&=&\dt
\end{eqnarray}
\end{subequations}
\section{Transitions from [1,2]}
\begin{subequations}
\begin{eqnarray}
\Pr([1,1]|[1,2])&=&\Pr(\text{success}|\er{Z})\\
\Pr([2,1]|[1,2])&=&\dt
\end{eqnarray}
\end{subequations}
\section{Initial distribution}
The initial distribution can be computed easily by noting the simple facts
that 
\begin{subequations}
\begin{eqnarray}
\er{Z}(\er{E}(\rho))&=&\er{E}(\rho)\\
\er{E}(\er{Z}(\rho))&=&\er{E}(\rho)\\
\er{Z}(\er{Z}(\rho))&=&\er{Z}(\rho)\\
\er{E}(\er{E}(\rho))&=&\er{E}(\rho).
\end{eqnarray}
\end{subequations}
Assuming that each $\CSIGN$ has two different failure modes -- control and
target failure -- we first calculate the probability of getting full erasures
for the desired pattern. Once that has been calculated, we calculate
the probability of getting $\Z$ erasures on the remaining unaffected qubits.
Following this procedure, one finds
\begin{subequations}
\begin{eqnarray}
\Pr([0,0])&=&(1-\e_l)^{14}\\
\Pr([0,1])&=&(1-\e_l)^7{7 \choose 1}\e_l(1-\e_l)^6\\
\Pr([1,0])&=&{7 \choose 1}\e_l(1-\e_l)^6(1-\e_l)^6\\
\Pr([1,1])&=&{7 \choose 1}\e_l(1-\e_l)^6{6 \choose 1}\e_l(1-\e_l)^5\\
\Pr([2,0])&=&{7 \choose 2}\e_l^2(1-\e_l)^5(1-\e_l)^5\\
\Pr([0,2])&=&(1-\e_l)^7{7 \choose 2}\e_l^2(1-\e_l)^5\\
\Pr([3,0])&=&\frac{4}{5}\left[{7 \choose 3}\e_l^3(1-\e_l)^4(1-\e_l)^4\right]\\
\Pr([0,3])&=&\frac{4}{5}\left[(1-\e_l)^7{7 \choose 3}\e_l^3(1-\e_l)^4\right]\\
\Pr([2,1])&=&\frac{4}{5}\left[{7 \choose 2}\e_l^2(1-\e_l)^5{5 \choose 1}\e_l(1-\e_l)^4\right]\\
\Pr([1,2])&=&\frac{4}{5}\left[{7 \choose 1}\e_l(1-\e_l)^6{6 \choose 2}\e_l^2(1-\e_l)^4\right],
\end{eqnarray}
\end{subequations}
and the probability of seeing an uncorrectable erasure is simply one minus
the sum of these probabilities.

%% file: thesis.bbl
\begin{thebibliography}{10}

\bibitem{aharonov}
{D}. {A}haronov and {M}. {B}en {O}r.
\newblock Fault-tolerant quantum computation with constant error rate.
\newblock 1999, quant-ph/9906129.

\bibitem{benioff}
{P}. {B}enioff.
\newblock The computer as a physical system: A microscopic quantum mechanical
  {H}amiltonian model of computers as represented by {T}uring machines.
\newblock {\em Journal of Statistical Physics}, 22(5):563--591, 1980.

\bibitem{steane:1996}
{A}.~{R}. {C}alderbank and {P}.~{W}. {S}hor.
\newblock {G}ood quantum error-correcting codes exist.
\newblock {\em Physical Review A}, 54:1098--1105, 1996.

\bibitem{chuang-yama:1995}
{I}. {C}huang and {Y}. {Y}amamoto.
\newblock {A} simple quantum computer.
\newblock {\em Physical Review A}, 52(5):3489--3496, 1995.

\bibitem{MAGMA}
Computational Algebra Group, School of Mathematics and Statistics, University
  of Sydney.
\newblock {\em {The Magma Computational Algebra System for Algebra, Number
  Theory and Geometry, Version 2.10}}, 2003.
\newblock \verb+(http://magma.maths.usyd.edu.au/magma/)+.

\bibitem{feynman}
{R}. {F}eynman.
\newblock {S}imulating physics with computers.
\newblock {\em {I}nternational {J}ournal of {T}heoretical {P}hysics}, 21:467,
  1982.

\bibitem{GAP4}
The GAP~Group.
\newblock {\em {GAP -- Groups, Algorithms, and Programming, Version 4.3}},
  2002.
\newblock \verb+(http://www.gap-system.org)+.

\bibitem{gottesman-thesis:1997}
{D}. {G}ottesman.
\newblock {\em {S}tabilizer {C}odes and {Q}uantum {E}rror {C}orrection}.
\newblock PhD thesis, California Institute of Technology, 1997.

\bibitem{gottesman-heisenberg:1998}
{D}. {G}ottesman.
\newblock {T}he {H}eisenberg {R}epresentation of {Q}uantum {C}omputers.
\newblock 1998, quant-ph/9807006.

\bibitem{gottesman:1998}
{D}. {G}ottesman.
\newblock {T}heory of fault-tolerant quantum computation.
\newblock {\em Physical Review A}, 57:127--137, 1998.

\bibitem{gottesman-chuang:1999}
{D}. Gottesman and {I}.~{L}. {C}huang.
\newblock {D}emonstrating the viability of universal quantum computation using
  teleportation and single-qubit operations.
\newblock {\em Nature}, 402:390--393, 1999.

\bibitem{grass-etal:1997}
{M}. {G}rassl, {T}. Beth, and {T}. {P}ellizzari.
\newblock {C}odes for the {Q}uantum {E}rasure {C}hannel.
\newblock {\em {P}hysical {R}eview {A}}, 56(1):33--38, 1997.

\bibitem{grassl-bch:1999}
{M}. {G}rassl and {T}h. Beth.
\newblock {Q}uantum {B}{C}{H} {C}odes.
\newblock 1999, quant-ph/9910060.

\bibitem{manny:2001}
{E}. {K}nill.
\newblock {L}inear {O}ptics {Q}uantum {C}omputation.
\newblock In {\em ITP Program on Quantum Information: Entanglement, Decoherence
  and Chaos}, pages I--V. (online), 2001.
\newblock http://online.itp.ucsb.edu/online/qinfo01/.

\bibitem{knill-bounds:2003}
{E}. {K}nill.
\newblock Bounds on the probability of success of postselected non-linear sign
  shifts implemented with linear optics.
\newblock 2003, quant-ph/0307015.

\bibitem{kl-cond:1997}
{E}. {K}nill and {R}. {L}aflamme.
\newblock Theory of quantum error-correcting codes.
\newblock {\em Physical Review A}, 55:900--911, 1997.

\bibitem{klm-thr:2000}
{E}. {K}nill, {R}. {L}aflamme, and {G}. {M}ilburn.
\newblock {T}hresholds for {L}inear {O}ptics {Q}uantum {C}omputation.
\newblock 2000, quant-ph/0006120.

\bibitem{klm:2001}
{E}. {K}nill, {R}. {L}aflamme, and {G}. {M}ilburn.
\newblock A scheme for efficient quantum computation with linear optics.
\newblock {\em Nature}, 409:46--52, January 2001.

\bibitem{klw}
{E}. {K}nill, {R}. {L}aflamme, and {W}. {Z}urek.
\newblock {T}hreshold {A}ccuracy for {Q}uantum {C}omputation.
\newblock 1996, quant-ph/9610011.

\bibitem{ralph:2003}
{A}.~{P}. {L}und, {T}.~{B}. {B}ell, and {T}.~{C}. {R}alph.
\newblock Comparison of linear optics quantum-computation control-sign gates
  with ancilla innefficiency and an improvement to functionality under these
  conditions.
\newblock {\em Physical Review A}, 68:022313, 2003.

\bibitem{lutkenhaus:1999}
{N}. {L}{\"u}tkenhaus, {J}. {C}alsamiglia, and {K}.-{A}. {S}uominen.
\newblock Bell measurement for teleportation.
\newblock {\em Physical Review A}, 59(5):3295--3300, May 1999.

\bibitem{boykin:1999}
{P}. {O}scar {B}oykin, {T}. {M}or, {M}. {P}ulver, {V}. {R}oychowdhury, and {F}.
  {V}atan.
\newblock {O}n {U}niversal and {F}ault-{T}olerant {Q}uantum {C}omputing.
\newblock In {\em {P}roceedings of the 40th {A}nnual {S}ymposium on
  {F}oundations of {C}omputer {S}cience}, page 486. IEEE {C}omp. {S}oc.
  {P}ress, 1999.

\bibitem{preskill}
{J}. {P}reskill.
\newblock {\em {I}ntroduction to {Q}uantum {C}omputation}, chapter
  {F}ault-tolerant quantum computation.
\newblock World Scientific, 1998, quant-ph/9712048.

\bibitem{zeilinger:1994}
{M}. {R}eck, {A}. {Z}eilinger, {H}.~{J}. {B}ernstein, and {P}. {B}ertani.
\newblock Experimental realization of any discrete unitary operator.
\newblock {\em Physical Review Letters}, 73(1):58--61, July 1994.

\bibitem{calderbank-etal:1998}
{A}. {R}obert {C}alderbank, {E}.~{M}. {R}ains, {P}.~{W}. {S}hor, and {N}.
  {J}.~{A}. {S}loane.
\newblock {Q}uantum {E}rror {C}orrection {V}ia {C}odes {O}ver {G}{F}(4).
\newblock {\em IEEE Transactions on Information Theory}, 44(4):1369 -- 1387,
  1998.

\bibitem{shor:1996}
{P}. {S}hor.
\newblock Fault-tolerant quantum computation.
\newblock In {\em 37th Symposium on Foundations of Computing}, pages 56--65.
  IEEE, IEEE Computer Society Press, 1996.

\bibitem{bank-shor:1996}
{A}.~{M}. {S}teane.
\newblock Multiple particle interference and quantum error correction.
\newblock {\em Proceedings of the Royal Society of London A}, 452:2551--2577,
  1996.

\bibitem{steane-filter:2002}
{A}.~{M}. {S}teane.
\newblock A fast fault-tolerant filter for quantum codewords.
\newblock 2002, quant-ph/0202036.

\bibitem{steane:2003}
{A}.~{M}. {S}teane.
\newblock {O}verhead and noise threshold of fault-tolerant quantum error
  correction.
\newblock {\em Physical Review A}, 68:042322, 2003.

\bibitem{steane-css-networks:2003}
{A}.~{M}. {S}teane and {B}. {I}bson.
\newblock {F}ault-{T}olerant {L}ogical {G}ate {N}etworks for {CSS} {C}odes.
\newblock 2003, quant-ph/0311014.

\bibitem{zalka:1996}
{C}. {Z}alka.
\newblock {T}hreshold {E}stimate for {F}ault {T}olerant {Q}uantum
  {C}omputation.
\newblock 1996, quant-ph/9612028.

\bibitem{zalka:personal}
{C}. {Z}alka.
\newblock Private communications.
\newblock 2003.

\bibitem{zlc:2000}
{X}. {Z}hou, {D}.~{W}. {L}eung, and {I}.~{L}. {C}huang.
\newblock Methodology for quantum logic gate construction.
\newblock {\em Physical Review A}, 62:052316, 2000.

\end{thebibliography}
